Division of Media, Communication, and Performing Arts

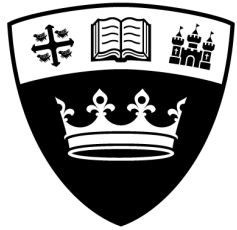

Matriculation Number: 22012759
Name of Supervisor: Dr. Isidoropaolo Casteltrione
Name of Degree Programme: BA (Hons) Media and Communications
Month and Year of Submission: April, 2025
Word Count: 9,844

# Deception Decoder: Proposing a Human-Focused Framework for Identifying AI-Generated Content on Social Media

*Submitted in partial fulfilment of the requirements for the degree of*
*BA (Hons) Media and Communications at Queen Margaret University*

# Acknowledgements

I would like to express my deepest gratitude to Dr. Isidoropaolo Casteltrione for their invaluable guidance and support as my supervisor throughout this research. Similarly, I must thank my programme leader, Dr. Taner Doğan, for fostering an intellectually engaging environment throughout the course, which directly inspired further my willingness to both achieve and excel.

I must also extend my utmost appreciation to my closest friends. Firstly, Stuart Imrie, for sharing with me his wealth of technical knowledge and understanding of machine learning processes. Secondly, Pelagia Demesticha, for lending me her social skills, without which the focus group session would have been impossible to hold together. Lastly, Kseniia Popova, for her genuine and continuous encouragement throughout the entire process.

Furthermore, I am indescribably grateful to Ella Phillips for her persistent and unwavering belief in my potential, which ultimately motivated me to pursue this programme of study; without you, this research would simply not exist. Likewise, I must thank my both of my loving parents, as well as my dear grandmother, for their unending patience, encouragement, well-intentioned criticisms, and most of all: support.

Finally, I acknowledge the faculty and staff of Queen Margaret University, Edinburgh, for providing the resources, environment, and inspiration, that made this project possible.

# Abstract

Generative AI (GenAI) poses a substantial threat to the integrity of information within the contemporary public sphere, which increasingly relies on social media platforms as intermediaries for news consumption. At present, most research efforts are directed toward automated and machine learning–based detection methods, despite growing concerns regarding false positives, social and political biases, and susceptibility to circumvention. This dissertation instead adopts a human-centred approach. It proposes the Deception Decoder—a multimodal, systematic, and topological framework designed to support general users in identifying AI-generated misinformation and disinformation across text, image, and video. The framework was developed through a comparative synthesis of existing models, supplemented by a content analysis of GenAI-video, and refined through a small-scale focus group session. While initial testing indicates promising improvements, further research is required to confirm its generalisability across user groups, and sustained effectiveness over time.

# Contents

# List of Figures

# List of Tables

# Chapter 1

# Introduction

Control over information—once mediated by legacy media institutions—has shifted dramatically to social media platforms such as TikTok, Instagram, Facebook, and X (formerly Twitter), which now act as arbiters of public discourse (Gillespie, 2018, pp. 1–9). These platforms now dominate how younger generations consume news, with over half of UK-based 18–24-year-olds citing such platforms as their primary news source (YouGov, 2024); similar global trends are observed (Watson, 2024). While social media is integral to the contemporary public sphere (Çela, 2015), it also relies upon opaque "black-box" algorithms (Reviglio and Agosti, 2020) that prioritise engagement, reinforcing 'filter bubbles' and 'echo chambers' (Kaplan, 2020; Terren and Borge, 2021), while amplifying disinformation—which demonstrably spreads faster than the truth (Vosoughi, Roy and Aral, 2018). These systems are further exploited by social bots, including emerging 'sleeper' variants, that mimic real users to distort discourse (Shao et al., 2018; Himelein-Wachowiak et al., 2021; Doshi et al., 2024).

The threat posed by this 'information disorder'—a term encompassing misinformation (unintentional error), disinformation (intentional deception), and malinformation (accurate but harmful content)—as defined by Wardle and Derakhshan (2017), is increasingly exacerbated by generative artificial intelligence (GenAI). Large Language Models (LLMs) such as ChatGPT (OpenAI, 2022) and image generation models such as Stable Diffusion (Esser et al., 2024) have become mainstream, while embedding themselves within fields such as journalism (Whittaker, 2019, p. 166; Newman and Cherubini, 2025). This is problematic, as LLMs struggle to consistently produce accurate information, even when this is their primary directive (Huang et al., 2023; Merken, 2023; Sherman and Kleinman, 2025). An increase in news content of this nature may increase the amount of misinformation present on social media when shared; this is without accounting for intentional malice. Recent elections and referendums were shaped by such mis/disinformation content (Allcott and Gentzkow, 2017; Henkel, 2021; West, 2024), as was the COVID-19 pandemic; which had severe ramifications toward the state of public health worldwide (Caceres et al., 2022; Sezer Kisa and Adnan Kisa, 2024). Genocide too; of the Rohingya in Myanmar (Amnesty International, 2022), and that which is 'plausible' in Gaza, Palestine (Amnesty International, 2024; Ali, Duggal and Salhani, 2024; OHCHR, 2024; Jones, 2025).

This study adopts a mixed-method approach. The literature review, in Chapter 2, explores GenAI mis/disinformation across text, image, and video; with a particular focus on the failure of automated detection tools and the potential of media and information literacy (MIL) interventions. Chapter 3 outlines the methodology, detailing a comparative analysis of existing frameworks, original content analysis of AI-generated video, and the use of participatory focus groups. The development, evaluation, and refinement of the proposed framework, including experimental results and thematic analysis are presented in Chapter 4. Finally, Chapter 5 concludes with a critical reflection on the framework's potential, limitations, and future adaptability.

In an environment where the distinction between true and false becomes increasingly blurred, this dissertation advocates for a dynamic, user-driven approach to detection in the battle against AI-generated mis/disinformation.

# Chapter 2

# Literature Review

The emergence of GenAI models has fundamentally altered the information ecosystem. With the increasing sophistication of tools capable of generating text, images, and video, misinformation and disinformation can now be created and disseminated at unprecedented scale and speed. This is particularly pertinent as the public becomes increasingly reliant on social media as a primary source of news (YouGov, 2024). This chapter reviews existing literature on threats introduced by GenAI; exploring the ways in which disinformation is spread, detected (or evades detection), and the importance of media and information literacy (MIL) in offering a more resilient, human-centred approach to mitigation. Ultimately, this chapter identifies critical gaps in technological and educational responses to GenAI-powered mis/disinformation, laying the groundwork for the development of a systematic, topological framework that prioritises human agency in the detection and evaluation of digital content.

## 2.1 GenAI-Text

The advancement of LLMs presents significant challenges to information integrity. This section examines existing research on GenAI-text, exploring the detection abilities of humans, as well as flaws within automated methods, while acknowledging the emerging risk of 'sleeper social bots'.

### 2.1.1 The Erosion of GenAI-Text Detectability

As of early 2025, there is a broad consensus among researchers that GenAI-text has eclipsed the point in which it becomes indistinguishable from that written by a human, both to the general public, as well as informed audiences. This has been confirmed relentlessly through research within the field (Kreps, McCain and Brundage, 2020; Clark et al., 2021; Spitale, Biller-Andorno and Germani, 2023; Zhou et al., 2023). For example, according to Clark et al. (2021), while text samples generated by earlier models such as GPT-2 (Radford et al., 2019), were identifiable in roughly 58% of cases, those produced by the more advanced GPT-3 (Brown et al., 2020) were only correctly identified as machine-generated approximately half the time. Consequently, the ability of an individual to correctly differentiate between human-written and GenAI-text is comparable to chance. With LLMs continuing to advance, traditional detection methods such as reliance on linguistic or self-contradictory errors produced within the model's output (Dou et al., 2021), have become increasingly less robust (Bashardoust, Feuerriegel and Shrestha, 2024). In the context of social media mis/disinformation, Spitale, Biller-Andorno and Germani (2023) and Zhou et al. (2023) corroborate these findings, demonstrating that LLM-generated COVID-19 disinformation is comparable to human-authored text in quality, and as such, difficult to discern. This suggests current state-of-the-art (SOTA) models are capable of creating imperceptible disinformation content suitable for dissemination on social media.

### 2.1.2 Willingness to Share Human vs. GenAI-Text

The potential impact of LLM-generated disinformation extends beyond creation; hinging on user engagement and sharing. Dupuis and Williams (2019) highlight the critical role of virality in amplifying disinformation campaigns on social media. Bashardoust, Feuerriegel and Shrestha (2024) found no significant difference in participants' willingness to share human-authored versus AI-generated 'fake news' excerpts, despite a 20% discrepancy in perceived veracity. Suggesting that even when users suspect content may be false, they are equally likely to propagate it.

### 2.1.3 Sleeper Social Bots

Doshi et al. (2024) identify the new phenomena of 'sleeper social bots', which utilise LLMs in order to dynamically disguise themselves as genuine users, with greater versatility than earlier counterparts (Shao et al., 2018). These 'sleeper' bots may more effectively engage in debate, and disseminate disinformation, or propaganda, by adapting contextually to user discussions in political contexts. The bots proved undetectable in the preliminary study conducted (Doshi et al., 2024), therefore presenting substantial risk toward increasing the number of sources pushing disinformation narratives on social media, while reducing their detectability.

### 2.1.4 Limitations of In-Model Mitigations

Efforts to embed safety mechanisms within LLMs (e.g., harmful prompt refusal), have proven easily circumventable, as illustrated through 'jailbreak' prompts (Wei, Haghtalab and Steinhardt, 2023; Spitale, Biller-Andorno and Germani, 2023) such as the infamous 'DAN' exploit (walkerspider, 2022) present within earlier versions of ChatGPT. These measures are only viable when LLMs are offered as one-to-many services by providers (Tang, Chuang and Hu, 2024), as is the case with ChatGPT or Gemini (Anil et al., 2023). When considering the rise of open-source, and local LLMs, matters are further complicated, as these models are both powerful, and customisable, which can include through the removal of such safeguards (Parthasarathy et al., 2024; k-mktr, 2025). Additionally, LLMs have demonstrated altered behaviour when they perceive surveillance from authority figures such as researchers (Greenblatt et al., 2024; Järviniemi and Hubinger, 2024; Park et al., 2024), introducing further doubt in the practical durability of these safety mechanisms.

### 2.1.5 Flaws in Automated Detection Solutions

Despite investment and adoption (Bort, 2024; GPTZero, 2024), automated, and machine learning (ML) based 'AI detectors' are demonstrably flawed. Liang et al. (2023) revealed significant bias against English as an Additional Language (EAL) writers, misclassifying their work as AI-generated 61% of the time—further corroborated by Perkins et al. (2024). Similarly, Rashidi et al. (2023) found LLM-detectors incorrectly flagged a considerable amount of scientific abstracts known to be author-written, as AI-generated. These detectors are circumventable through adversarial techniques such as altering prompts (Zhang et al., 2024), introducing spelling errors (Shushanta Pudasaini et al., 2025), additional punctuation and spaces (Cai and Cui, 2023), or otherwise altering linguistic characteristics. Therefore, they are not fit for purpose as a sole method concerning the detection of mis/disinformation on social media. Similarly, 'watermarking' and 'retrieval-based' approaches are also likely to become obsolete as LLMs continue to evolve (Kirchenbauer et al., 2023; Tang, Chuang and Hu, 2024), and perhaps, due to the availability of SOTA open-source models such as Deepseek-R1 (DeepSeek-AI et al., 2025), already are.

Alternative approaches, such as those which posit that LLMs themselves be installed within automated fact-checking systems (Quelle and Bovet, 2024), raise substantial concerns due to the inherent political and social biases within the models themselves (Zhao et al., 2023; Adewumi et al., 2024; Bai et al., 2024; Bang et al., 2024; Ceron et al., 2024; Rozado, 2024; Shin et al., 2024; Wright et al., 2024). This particularly proves troubling as it concerns perspectives from the global south, and geopolitical bias (P et al., 2024).

Reliance on such demonstrably flawed systems risks not just discrediting genuine content, as far as academics, or journalists who increasingly work collaboratively with LLMs (Whittaker, 2019, p. 166; Adami, Kahn and Suárez, 2025), but also risks reinforcing existing cognitive bias. Zhu et al. (2024) identify that participants universally preferred text labelled as human-written, even when this was not the case. Therefore, labelling social media posts in such a manner may inadvertently increase the legitimacy of LLM-generated disinformation content, which may employ mitigations to evade detection, while delegitimising those who use such tools collaboratively—or not at all, in the case of EAL writers. Regardless, an over-reliance on such systems leading to this outcome could both severely damage reputation, and instil false trust mis/disinformation content.

### 2.1.6 MIL and Human-Focused Detection

MIL intervention at the human level is posited as a potential mitigation by several authors of the reviewed works (Kreps, McCain and Brundage, 2020; Zhou et al., 2023; Bashardoust, Feuerriegel and Shrestha, 2024). Despite this, there exists only one comprehensive, and human-focused framework, which codes the errors and anomalies typically present within AI-generated text, Dou et al.'s (2021) 'Scarecrow'. However, due to the advancement of LLMs since the original publication, the model proposed is now obsolete in several categories; necessitating either a revision or adaption for use in the contemporary context.

### 2.1.7 Conclusion

The use of LLMs to generate disinformation at scale appears inevitable—just as likely is the eventual defeat of current ML-based detection tools (Tang, Chuang and Hu, 2024). The present research has mainly focused on in-model mitigations, and external detection utilising ML, or classifiers, which are circumventable by a determined, malicious entity, such as someone wishing to disseminate a campaign of disinformation on social media. Emphasising the need for a manual line of defence employing critical thinking.

## 2.2 GenAI-Images

Mis/disinformation in the social media context is significantly challenged by the rapid advancement of GenAI-image/video. This section examines existing research on GenAI visual content, focusing on the limitations of current frameworks, identifying key artifacts, and exploring potential detection methods.

### 2.2.1 GenAI-Images and Mis/Disinformation

Countless GenAI-images, many of which are politically motivated, circulate online (Devlin and Cheetham, 2023; Vincent, 2023; Wendling, 2024; Wilson, 2025). Notably, OpenAI (2024b) has published extensive reports on the misuse of their models. Furthermore, these tools are becoming more accessible. Mehta et al. (2023) found that novice users can create convincing 'deepfakes'—which are falsified likenesses of individuals—with adequate resources and time. Confirming prior concerns about unskilled and widespread use (Hwang et al., 2021; Helmus, 2022). 'Deepfakes' have seen widespread use in scams employing the likenesses of famous individuals (ITV News, 2023; Chen and Magramo, 2024; Price, 2024; Gozzi, 2025), as well as for politically manipulative purposes (Momeni, 2024). While the falsification of images and video is hardly new, the scale at which this may be conducted, as well as by whom, is. Many such tools are now freely available, open-source, and deployable on consumer-level hardware (k-mktr, 2025).

### 2.2.2 Human Detection of GenAI-Images

Early research highlighted an inability among individuals to reliably distinguish between real and AI-generated images. Kas et al. (2020) demonstrated that participants correctly identified genuine faces at a rate of 81%, compared to only 61% for GAN-generated faces; with some generated images achieving 100% misclassification rates. Bray, Johnson, and Kleinberg (2023) corroborated these findings, reporting an overall accuracy rate of 62% in differentiating between real and GAN-generated images, even when participants received textual advice on identifying key artifacts—abnormalities within such media. Suggesting that current intervention strategies are insufficient, potentially due to the nature of the guidance provided; the incorporation of image or video examples, rather than text-only intervention, would prove more effective. Furthermore, both studies noted a tendency for participants to overestimate their accuracy.

Similarly, Lu et al. (2023), comprehensively benchmarked human performance in the accurate detection of outputs from both the aforementioned GANs, as well as contemporary 'diffusion' models through their 'HPBench' experiment spanning several categories: 'multiperson', 'landscape', 'man', 'woman', 'record', 'plant', 'object', and 'animal'. Such diversity is integral in understanding the performance of image generation models in a variety of circumstances. The results confirm that humans struggle to reliably determine the legitimacy of images, with an overall misclassification rate nearing 39%. Diffusion-generated images—hereafter referred to as GenAI-images—were incorrectly classified as real an average of 44% of the time, although this varied by category. Notably, images featuring humans were more often correctly identified than those depicting objects – a crucial consideration given the risks associated with impersonation in disinformation campaigns (Momeni, 2024). Conversely, real images were only identified as such 70% of the time, which the authors suspect is symptomatic of the growing distrust in genuine media. Overall, an 11% gap between the human ability to correctly identify real images, and incorrectly identify false images was identified, with Lu et al. (2023) predicting that this shall narrow as generative models continue to improve.

### 2.2.3 In-Model and External Detection

Recognising the potential for misuse, developers are exploring both internal mitigations within generative models and external detection methods. OpenAI's (2024a; Betker et al., 2023) DALL·E 3, for example, is trained to avoid generating images of public figures. These safeguards are only effective when the developer has full control of the model's use and output. This is not the case with open-source solutions such as Stable Diffusion 3 (Esser et al., 2024), or Black Forest Labs' (2024) FLUX.1-dev, which may be further modified with specialised adapters or 'LoRAs' (i.e. those found on Civitai, 2025). Furthermore, methods such as fingerprinting, or watermarking GenAI-images during the generation process, so that they may later be recognisable using purpose-built detection systems, have also been explored (Liao et al., 2022; Fernandez et al., 2023; Liu et al., 2023; Sinitsa and Fried, 2023). While more robust than 'passive' methods used in commercial detectors (Guo et al., 2024), they remain vulnerable to adversarial attacks (Li et al., 2023). Furthermore, evidence suggests that watermarks may be injected into genuine images (Saberi et al., 2023), exposing the risk of an over-reliance on such methods degrading trust in all photographic media.

### 2.2.4 Human vs. Machine Detector Performance

Ha et al. (2024) compared the performance of humans and ML-based detectors in identifying AI-generated artworks, while testing resilience against common adversarial attacks (e.g., JPEG compression, Gaussian noise). The study employed three human cohorts: general crowd-workers, professional artists, and expert identifiers, alongside three ML-based detectors. Results indicated challenges for humans, with the general cohort achieving only 59% accuracy, while alterations to GenAI-images reduced the efficacy of ML-based solutions.

### 2.2.5 Artifacts Within GenAI-Images

Mathys, Willi, and Meier (2024) conducted a comprehensive content analysis of GenAI-images to provide visual guidance for identifying synthetic media. Their findings highlight specific challenges of GenAI-images accurately depicting: groups of people, background elements, or specialist objects such as vehicles, resulting in an increased presence of artifacts in such scenes. For objects in particular, expert knowledge of the subject matter is deemed crucial for authentication.

### 2.2.6 Overview of GenAI-Video

'Deepfakes' already represent a significant, and well-studied aspect of digital disinformation theory, with several detection strategies already devised (Li, Chang and Lyu, 2018; Rössler et al., 2018; Matern, Riess and Stamminger, 2019; Mirsky and Lee, 2021). However, these are distinct from video generation systems as they concern this research. This GenAI-video lacks research regarding the investigation of similar strategies, possibly due to high-quality generators being a recent development.

These systems are architecturally similar to the previously discussed 'diffusion' image generators in the case of OpenAI's SORA (Brooks et al., 2024), Tencent's HunyuanVideo (Kong et al., 2024), Alibaba's Wan (WanTeam et al., 2025), and others (Peng et al., 2025). Detection methods for this type of content likely mirror those identified in Mathys, Willi and Meier's (2024) content analysis.

### 2.2.7 Conclusion

The advancement of GenAI systems capable of producing realistic visual content presents challenges to both ML and human detection capabilities. Even informed individuals struggle to reliably distinguish between real and GenAI-images, and ML-based detectors—though promising in controlled settings—remain vulnerable to defeat. Fingerprinting and watermarking, while conceptually robust, face issues of circumvention and trust when deployed at scale. The convergence of open-source GenAI models, low barriers to entry, and increasingly sophisticated video generators, suggest a near future in which AI-generated media becomes indistinguishable from authentic content. These developments highlight the need for detection frameworks that are technically adaptable but also human-centred, incorporating MIL and critical engagement.

## 2.3 MIL Intervention

The increasing sophistication of GenAI presents a significant challenge to information integrity. Kaplan (2020) highlights the crucial role of education in empowering citizens to critically evaluate online content, arguing for the broad integration of AI awareness into curricula. Many existing works emphasise MIL interventions as key tools in combating risks posed by AI-generated misinformation and disinformation (Kreps, McCain and Brundage, 2020; Hwang et al., 2021; Helmus, 2022; Zhou et al., 2023; Bashardoust, Feuerriegel and Shrestha, 2024; Mathys, Willi and Meier, 2024). This section examines current frameworks addressing information disorder (Wardle and Derakhshan, 2017), with a particular focus on their capacity to account for the evolving threat of GenAI.

### 2.3.1 The Efficacy of MIL Interventions

Adjin-Tettey (2022) found that even a brief, two-hour training session significantly improved participants' accuracy in identifying disinformation—increasing from 46% correct identification pre-training to 73% post-training. Similarly, Moore and Hancock (2022) observed similar improvements in an older adult population following a self-directed, hour-long interactive module focusing on text, image, and video misinformation. Accuracy rates improved from 64% to 85% post-intervention, with participants also demonstrating increased adoption of strategies for identifying misinformation.

### 2.3.2 Limitations of Existing MIL Frameworks

Tiernan et al. (2023) identify a critical gap in current frameworks: their limited capacity to address threats posed by GenAI. Their analysis of eleven MIL frameworks reveals an overarching inflexibility stemming from their traditional 'report-style' or published nature. This rigidity hinders the rapid iteration necessary to address the threats posed by a constantly evolving adversary, such as GenAI. The authors advocate for adopting inspiration from 'agile' principles (Beck et al., 2001)—common in software design—within framework creation, emphasising, individuals and interactions over processes and tool; or rather frameworks developed with direct input from concerned stakeholders. An online-first, adaptable approach, coupled with crowdsourced maintenance involving educators and students, is proposed as a means to ensure continued relevance. Furthermore, Shalevska's (2024) analysis of MIL handbooks reveals a significant oversight regarding GenAI, the surveyed materials lacked comprehensive definitions or explanations. While acknowledging the problem of 'fake news', none addressed the specific role of AI in generating and disseminating misinformation.

### 2.3.3 Conclusion

MIL remains one of the most promising human-centred approaches to countering the threats posed by GenAI, however, current frameworks exhibit significant limitations. The reviewed literature makes clear that targeted interventions can improve users' ability to detect disinformation; these efforts must be embedded within user-focused, and regularly updated educational structures. Incorporating AI awareness, detection strategies, and critical evaluation skills into frameworks is no longer optional but essential. Similarly, frameworks must evolve from static to dynamic in implementation, with direct stakeholder involvement in the design process.

## 2.4 Research Gap

Most of the works surveyed explore automated solutions for detecting AI-generated mis/disinformation, which is problematic for the issues of bias, and inaccuracies as outlined. Furthermore, a notable gap in the literature is the overwhelming reliance on quantitative over qualitative methods. While quantitative approaches offer objective measures of trends, they overlook nuance that perhaps would be uncovered through a qualitative approach. Highlighting a potential loss of depth in understanding user behaviours and perceptions. Furthermore, most research focuses only on the detection of GenAI-text or GenAI-images, and not both together in a mis/disinformation context.

This necessitates studies that combine quantitative and qualitative approaches to better understand trends and human perception. Developing robust detection systems and promoting MIL is crucial for better mitigating the impact of mis/disinformation. The myriad of ML-based solutions all present flaws in their own right; despite this, a human-focused, comprehensive, multimodal, and modular detection framework has not yet been created.

This study aims to develop a systematic, topological, and human-focused framework to successfully aid the detection of GenAI-text, GenAI-images, and GenAI-video across social media. Creating such a framework would pose an alternative to the strong research focus on automated, ML-based approaches; while delivering on the need for MIL tools to directly challenge the risks posed by AI-generated mis/disinformation.

# Chapter 3

# Methodology

This chapter outlines the research design, methods of data collection, analysis, and ethical considerations employed. Subsequently, this study aimed to create an adaptable, human-centred, systematic framework for detecting AI-generated mis/disinformation on social media.

## Research Questions

- **RQ1:** What elements are present within existing frameworks, and how do they conceptualise and account for GenAI-driven mis/disinformation, including the roles of 'sleeper' social bots, content types (e.g., text, image, video), and end-user support for detection and mitigation?

- **RQ2:** What artifacts, anomalies, or inconsistencies can be identified as either in common with image counterparts, or unique to GenAI-videos, and how might these inform detection strategies?

- **RQ3:** What visual, linguistic, and contextual cues do individuals rely on to detect AI-generated mis/disinformation, and how can human-focused frameworks enhance these detection capabilities?

- **RQ4:** How can participatory research methods inform the development and refinement of human-centred frameworks for detecting AI-generated mis/disinformation?

## 3.1 Research Design

A mixed-method research approach was adopted, combining both qualitative and quantitative techniques. This research design is highly applicable for those seeking to broach real-world issues such as AI-generated mis/disinformation, through the development of instruments or interventions (Tashakkori and Teddlie, 2003; Creswell and Plano Clark, 2018); such as the developed '*Deception Decoder* framework.

This integration also provided methodological and data triangulation, enhancing credibility by drawing from distinct methods and data sources; including academic frameworks, synthetic content, and participant feedback. As a result of the dual aims of the study, the research is structured accordingly:

1. **Designing the Preliminary Framework** – This was informed by the comparative framework analysis and quantitative content analysis of GenAI-videos. Then synthesising existing, fragmented, works into a unified model, with expansion to cover GenAI-video.

2. **Finalising the Framework** – Focus groups were employed to evaluate pre- and post-framework intervention detection accuracy through an experiment, as well as to gather alternate perspectives and suggestions for improvement of the preliminary model (Tiernan et al. 2023), this then formed the final draft of the *Deception Decoder*.

## 3.2 Designing the Preliminary Framework

### 3.2.1 Comparative Framework Analysis (RQ1)

A comparative analysis was conducted to evaluate four key frameworks, purposively selected for their relevance to information disorder and GenAI, with each representing a different aspect of the mis/disinformation landscape:

- **Wardle and Derakhshan (2017)** – foundational definitions of information disorder.

- **Dou et al. (2021)** – GenAI-text detection through error categorisation.

- **Mathys, Willi and Meier (2024)** – artifact-based analysis, and coding of GenAI-images.

- **Shalevska (2024)** – media literacy modules incorporating GenAI and social bots.

**These frameworks were systematically evaluated using a matrix based on the following predefined criteria:**

- Does it address information disorder (mis/disinformation)?

- Does it address social bots (including 'sleepers')?

- Does it address AI/GenAI, and in which modalities (e.g., text/image/video)?

- Who is the target audience (e.g., researchers vs. end-users)?

- Contemporary relevance and limitations

This comparative process employed the 'Framework Method' (Ritchie and Spencer, 1994), a structured approach to qualitative analysis involving familiarisation, indexing, charting, and interpretation. A matrix was then developed to systematically evaluate each framework against predefined thematic criteria, allowing for a visual, and conceptual comparison of their key features. Each framework was treated as a distinct conceptual case, analysed holistically in alignment with case-orientated principles (Ragin, 2014, pp. 51–52), addressing the frameworks as integrated wholes rather than as isolated variables. This approach revealed that while each framework contributed valuable elements, none alone provided an accessible, multimodal human-focused solution. A structured comparison informed the synthesis of the initial three-layer preliminary *Deception Decoder* framework, hereafter referred to as the 'preliminary framework'. Resultantly, the following elements were devised: Source (trust classification), Content (GenAI-text/image/video artifact recognition), and Motive (intent categorisation).

### 3.2.2 Content Analysis of GenAI-Video

Both GenAI-text and GenAI-images have seen substantial research focus, and thus, were incorporated into the preliminary framework through the synthesis of existing frameworks. Contrastingly, GenAI-video remains under-researched. To enable the methodologically sound inclusion of this modality within the *Deception Decoder* an analysis of this content was deemed necessary. Audio was excluded from the content analysis, as SOTA video generators do not currently produce accompanying audio outputs; making audio-based detection outside the scope of this study.

A problem-driven quantitative content analysis (Neuendorf, 2017, pp. 44-45; Krippendorff, 2018, pp. 384-387) was conducted on 20 GenAI-videos (see: Bowman Kerbage, 2025a), as to identify both shared—with GenAI-images—and emergent artifacts within this modality. These were generated using Tencent's 'HunyuanVideo' model (Kong et al., 2024), with 'Boreal-HL' LoRA (kudzueye, 2025), an experimental style adaptor that enhances visual realism. This model was selected for its open-source nature, and high visual realism, representative of SOTA tools currently available for potential misuse.

As video generators are architecturally similar to their image generation counterparts (see: Section 2.2.6), Mathys, Willi, and Meier's (2024) GenAI-image artifact taxonomy was utilised, allowing for direct comparison to the results of this work (Neuendorf, 2017, p. 150). Initial codes included:

- **Physical**: Violations of physics like incorrect lighting, shadows, reflections, or floating objects without support.

- **Geometrical**: Distorted shapes, incorrect perspective, or misaligned patterns that break spatial realism.

- **Human Anatomy**: Unnatural features in faces, hands, or body proportions that look 'wrong' to human eyes.

- **Distortions**: Visual noise, blurring, odd colours, or unintended styles that reduce image clarity and realism.

- **Text**: Scrambled, unreadable, or inconsistent text that fails to resemble real writing.

- **Semantic**: Logically implausible scenes or object relationships that do not match real-world expectations.

Additionally, two emergent codes were identified inductively through repeat empirical examinations of the content:

- **Continuity Artifacts**: Visually unnatural changes in objects over time (e.g. items morphing, merging, or vanishing abruptly).

- **Animation Artifacts**: Implausible or unrealistic motion (e.g. floaty walking, wheels that do not spin properly).

## 3.3 Preliminary Framework Evaluation and Iteration

Focus groups, as a method of participatory research, present an ideal platform enabling the co-construction of the framework through direct stakeholder input, as recommended in literature (Tiernan et al., 2023). The session involved several phases including both experiments, to gain an exploratory understanding of framework efficacy, and group discussions (Flick, 2014). These were chosen to explore in-depth perspectives of the cohort, as well as suggestions for improvement of the preliminary framework. For more details, refer to Appendix A.

### 3.3.1 Focus Group Experiment (RQ3)

The purpose of this experiment was to gain an exploratory understanding as to whether detection accuracy could be improved through the introduction of the preliminary framework. A within-subjects approach (Field, 2024, p. 1070), employing a pre/post structure was utilised: participants were presented with a series of social media posts, half of which were real mis/disinformation sourced from 'X', while the remainder were AI-generated (for further details, see: Appendix A.2). Participants were asked to determine whether each was AI-generated or real mis/disinformation and record their decision on an anonymised answer sheet. Following this, the framework was introduced and the same task repeated. Detection accuracy scores were recorded across three categories: text-only, text+image, and text+video.

To assess improvement, descriptive statistics were employed: by category to better understand how differences in modality affect detection, as well as by participant. Additionally, effect size was calculated using Cohen's $d$, measured with reference values ($small$=0.2, $medium$=0.5, $large$=0.8), as to quantify the magnitude of any observed improvement (Cohen, 1988, pp. 20–27). Effect size was prioritised over statistical significance due to the small sample size ($n$=4) resulting in a lack of power, which could therefore produce misleading results; whereas effect size is not dependant on sample quantity, and provides a more meaningful measure of practical significance (Field, 2024, p. 129).

## 3.3.2 Thematic Analysis of Focus Group Session (RQ3/RQ4)

Participant feedback was recorded, and then transcribed using WhisperX (Bain et al., 2023). The transcript was anonymised, and analysed thematically using Braun and Clarke's (2006) six-step process. This approach involved coding participant feedback to identify recurring themes across the focus group session, which would later directly inform the iterative refinement of the framework, particularly in adjusting it to align with their real-world detection strategies, and concerns of usability.

**Current Detection Strategies (RQ4)**

Firstly, participants were asked of their current concerns, and strategies for detection, regarding AI-generated content on social media – as to answer RQ3. Themes emerged across four key areas:

- Visual Anomalies in GenAI-images & GenAI-videos
- Linguistic Red Flags
- Context as a Detection Strategy
- Concerns About AI-Generated Content

However, as the topics within these findings were represented already within the preliminary framework, no action was taken.

**Suggestions for Framework Improvement (RQ4)**

Additionally, participants discussed possible refinements to the proposed model, as to address their concerns of usability and accessibility. The following themes emerged, answering RQ4:

- Better Personalisation & Training
- Providing Practical Guidance
- Incorporating a Philosophical & Educational Perspective
- Improving Usability & Clarity

These findings were then mapped to framework components and directly informed revisions. The participatory nature of this feedback process aligns with Tiernan et al. (2023), who highlight the importance of stakeholder input in designing practical interventions.

**Three core changes were made based on focus group findings:**

- Implemented a 'Why This Matters' section to provide context and usage guidelines

- A 'Three Strikes' rule was added to encourage actionable detection

- Technical terminology was simplified

The revised version of the framework was uploaded to GitHub (see: Bowman Kerbage, 2025b) to allow for ongoing community contribution and refinement, in alignment with prior recommendations from research (Tiernan et al., 2023).

### 3.3.3 Sampling

This study utilises a mix of purposive and convenience sampling due to time and feasibility constraints. Notably, this cohort was engaged in their own research during the time-frame of this research, and due to the demanding nature of the focus group session (containing both experiments and comprehensive discussion), several last-minute withdrawals occurred. Regardless, the small sample size ($n=4$), is acceptable, but only due to the exploratory nature of this research. Participants were known to be final-year students in either Media and Communications, or Film and Media courses at Queen Margaret University; programmes which cover misinformation, disinformation, and 'fake news' across several modules. Existing familiarity with the topic was pertinent, as this cohort's responses would directly influence the framework proposed through discussion within the focus group session.

### 3.3.4 Limitations

Due to the explorative nature of this research, limitations include a small sample size ($n=4$), homogeneity of participants (final-year media students), and reliance on a single video generation model. However, as this study focused on the design of the framework itself, rather than its validation outwith proof-of-concept, or generalisability of findings, these limitations aligned with the project's scope and focus. Additionally, as the focus group cohort shares pre-existing knowledge of mis/disinformation, 'fake news,' and media literacy concepts, their responses may have introduced biases within the data. As a result, the insights gathered reflect an informed viewpoint rather than those of the general population.

### 3.3.5 Ethical Considerations

Ethical approval for this research has been obtained from Queen Margaret University's Research Ethics Committee, and Informed consent was gained from all participants. The session was audio recorded for transcription, which was carried out immediately using WhisperX (Bain et al., 2023). Once transcribed, the original audio recording was erased, while the transcript was anonymised. All data utilised in this research will be securely stored for a period of 90 days commencing April 1st 2025, after which it will be permanently deleted, complying with the UK GDPR, and Data Protection Act 2018 (legislation.gov.uk, 2016; 2018).

# Chapter 4

# Results

This section begins by presenting the findings from the comparative analysis of four frameworks addressing information disorder (Wardle and Derakhshan, 2017) in the context of GenAI and social bots. Following this, insights from the content analysis of GenAI-video data are discussed alongside key themes emerging from the focus group experiments, and discussions with end-users about their experiences navigating disinformation online, culminating in the proposal of a systematic, topological, and user-focused detection framework for AI-generated disinformation on social media.

## 4.1 Designing the Preliminary Framework

Please turn over...

### 4.1.1 Comparative Analysis of Existing Frameworks (RQ1)

A comparative analysis was undertaken to addressing and investigate RQ1, through the identification of the strengths and weaknesses present within current frameworks.

| Framework | Wardle & Derakhshan (2017) | Dou et al. (2021) (Scarecrow) | Mathys, Willi, and Meier (2024) | Shalevska (2024) |
|---|---|---|---|---|
| Purpose/ Scope | Defines information disorder (types, phases, elements) | Categorises errors in GenAI-text | Identifies visual artifacts in GenAI-images | Develops educational modules for AI literacy/awareness |
| AI Focus | Mentions AI's role in automation and bias, lacks GenAI analysis | Focuses on GenAI-text errors, not AI-driven dissemination (e.g., social bots) | Analyses GenAI-image artifacts | Focuses on detecting GenAI-driven mis/disinformation |
| Target Audience | Policymakers, researchers, practitioners | Researchers, academics studying LLMs | Researchers, educators, policy-makers | Educators designing curricula for GenAI literacy or awareness |
| Relevance Today | Foundational but outdated for modern GenAI (e.g., diffusion generators) | Still useful for text error analysis, but outdated with newer LLMs | Highly relevant for detecting GenAI-images | Critical for media literacy around GenAI content |
| Social Bots Addressed? | Partially: Acknowledges bot-driven amplification, lacks focus on advanced GenAI content distribution | No: Focuses on text flaws, not bot-driven disinformation | No: Focuses on image artifacts, not bot networks | Yes: Covers bot networks, social media manipulation, and 'sleeper' bots |
| Key Limitations | No GenAI integration (e.g., LLMs, text-to-image/video); lacks end-user focus | Limited relevance to current LLMs; does not address GenAI-images/videos | Largely framed for an informed or aware audience not general users | Lacks visual examples; may need adaptation for different contexts |

Table 4.1: Comparative matrix assessing four key mis/disinformation and GenAI detection frameworks according to the aforementioned criteria.

The table above describes four key frameworks targeting mis/disinformation or GenAI content detection. While each framework individually contributes valuable perspectives, such as foundational definitions, educational focus, and identification strategies, none simultaneously address all investigated competencies. This highlights the fragmented nature of current approaches, particularly in terms of addressing social bots, multimodal (across text, image, and video) AI-generated mis/disinformation, from a general or end-user perspective; which has been overlooked.

## 4.1.2 Quantitative Content Analysis of GenAI-Video (RQ2)

This content analysis of 20 GenAI-videos (Bowman Kerbage, 2025a) seeks to answer RQ2, through deductively employing codes from Mathys, Willi, and Meier's (2024) framework. Audio was excluded from this analysis, as current video generators do not create audio alongside visual output.

| Video | Physical | Geometrical | Human Anatomy | Distortions | Text | Semantic | *Animation* | *Continuity* | Total Errors |
|---|---|---|---|---|---|---|---|---|---|
| Video 1 |  |  |  |  |  |  |  |  | 5 |
| Video 2 |  |  |  |  |  |  |  |  | 3 |
| Video 3 |  |  |  |  |  |  |  |  | 4 |
| Video 4 |  |  |  |  |  |  |  |  | 6 |
| Video 5 |  |  |  |  |  |  |  |  | 0 |
| Video 6 |  |  |  |  |  |  |  |  | 5 |
| Video 7 |  |  |  |  |  |  |  |  | 7 |
| Video 8 |  |  |  |  |  |  |  |  | 7 |
| Video 9 |  |  |  |  |  |  |  |  | 8 |
| Video 10 |  |  |  |  |  |  |  |  | 2 |
| Video 11 |  |  |  |  |  |  |  |  | 6 |
| Video 12 |  |  |  |  |  |  |  |  | 6 |
| Video 13 |  |  |  |  |  |  |  |  | 4 |
| Video 14 |  |  |  |  |  |  |  |  | 6 |
| Video 15 |  |  |  |  |  |  |  |  | 4 |
| Video 16 |  |  |  |  |  |  |  |  | 4 |
| Video 17 |  |  |  |  |  |  |  |  | 6 |
| Video 18 |  |  |  |  |  |  |  |  | 4 |
| Video 19 |  |  |  |  |  |  |  |  | 4 |
| Video 20 |  |  |  |  |  |  |  |  | 6 |
| Per Artifact Type | Physical | Geometrical | Human Anatomy | Distortions | Text | Semantic | *Animation* | *Continuity* | Overall Mean |
| Total Occurrence | 8 | 8 | 17 | 14 | 11 | 17 | 10 | 12 | 4.85 |
| Occurrence Rate | 40.00% | 40.00% | 85.00% | 70.00% | 55.00% | 85.00% | 50.00% | 60.00% | 60.63% |

Table 4.2: Matrix visualisation of the content analysis, including the identified *supplemental codes*. Red cells indicate an artifact of this nature was identified within the analysed video.

Of the 20 videos analysed, a mean artifact occurrence rate of 60% (4.85/8) was observed. Anomalies in the categories of 'Human-anatomy' and 'Semantic' were the most prevalent, both with an occurrence rate of 85%. Artifacts concerning 'Distortion', and 'Text' were also frequently observed at the rate of 70% and 55% respectively. The lowest occurring artifact types, 'Physical', and 'Geometrical', both had a prevalence of 40%. These factors suggest that GenAI-video exhibits all of the same flaws as can be observed in GenAI-images, thus informing RQ2.

**Identified Supplemental Codes**

Additionally, through the content analysis, two supplemental categories exemplifying the detectable characteristics unique within GenAI-video were identified through visual observation, presenting an answer to RQ2:

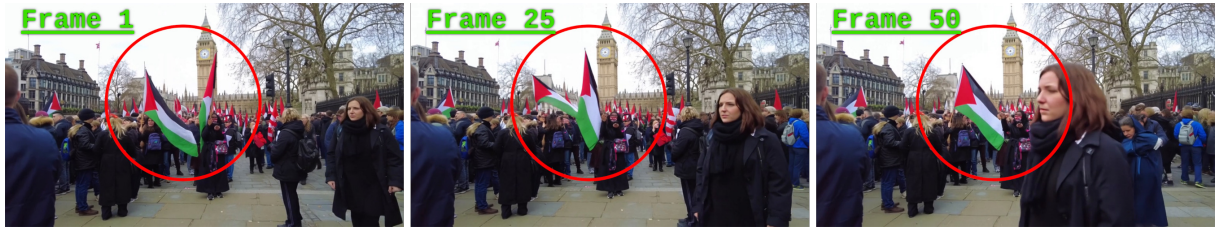

Figure 4.1: An example of the 'Continuity' artifact selected from Video 20 (Bowman Kerbage, 2025a).

- **Continuity:** These artifacts occur when the video depicts objects which experience visually unnatural transformations. For example, items changing shape or size as the scene progresses, merging with other elements, or entirely vanishing with no plausible explanation.

- **Animation:** Movement which displays implausible characteristics, such as appearing 'floaty', or otherwise failing to represent reality. This may concern depictions such as those of people with irregular walking techniques, and vehicles which display inconsistent patterns on in-motion components (e.g., wheels).

### 4.1.3 Discussion and Synthesis

The comparative analysis of existing frameworks reveals distinct strengths and notable limitations, many of which shaped the development of the preliminary detection framework presented in this study.

Figure 4.2: The three types of information disorder as presented by Wardle and Derakhshan (2017).

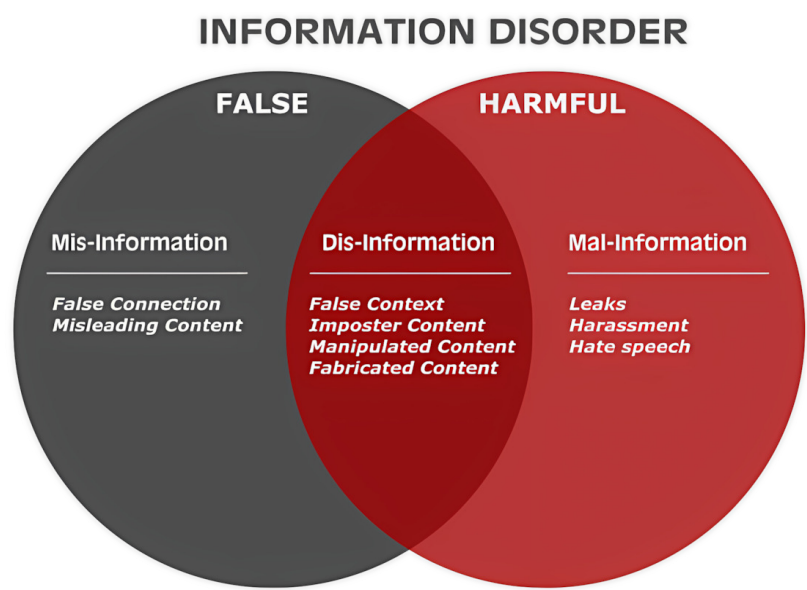

Wardle and Derakhshan's (2017) framework provides a comprehensive array of foundational definitions for information disorder, clarifying misinformation, disinformation, and malinformation as distinct concepts. This directly informed the 'Motive Matrix' within the proposed framework, as to help users distinguish between honest mistakes and deliberate acts of deception. Their treatment of the 'actors' within information disorder, as outlined in the 'agent' component, also influenced the design of the 'Source' layer, shaping the classification system into 'Trusted', 'Cautionary', and 'Untrusted' categories. However, the framework predates the rise of GenAI, and thus understandably lacks engagement with current threats such as AI-generated media or the emergence of 'sleeper' bots (Doshi et al., 2024). Limiting its relevance to present-day challenges, though still providing essential structural definitions that support more contemporary models.

Figure 4.3: The three elements of information disorder as presented by Wardle and Derakhshan (2017).

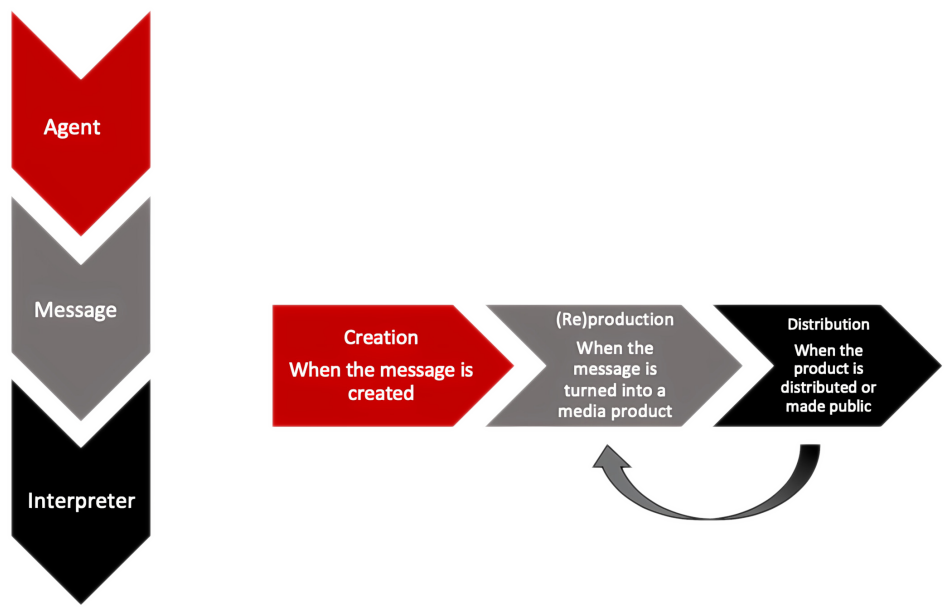

Figure 4.4: An overview of Wardle and Derakhshan's (2017) '*Information Disorder*' as synthesised within the *Deception Decoder* framework's <u>Source and Motive</u> nodes.

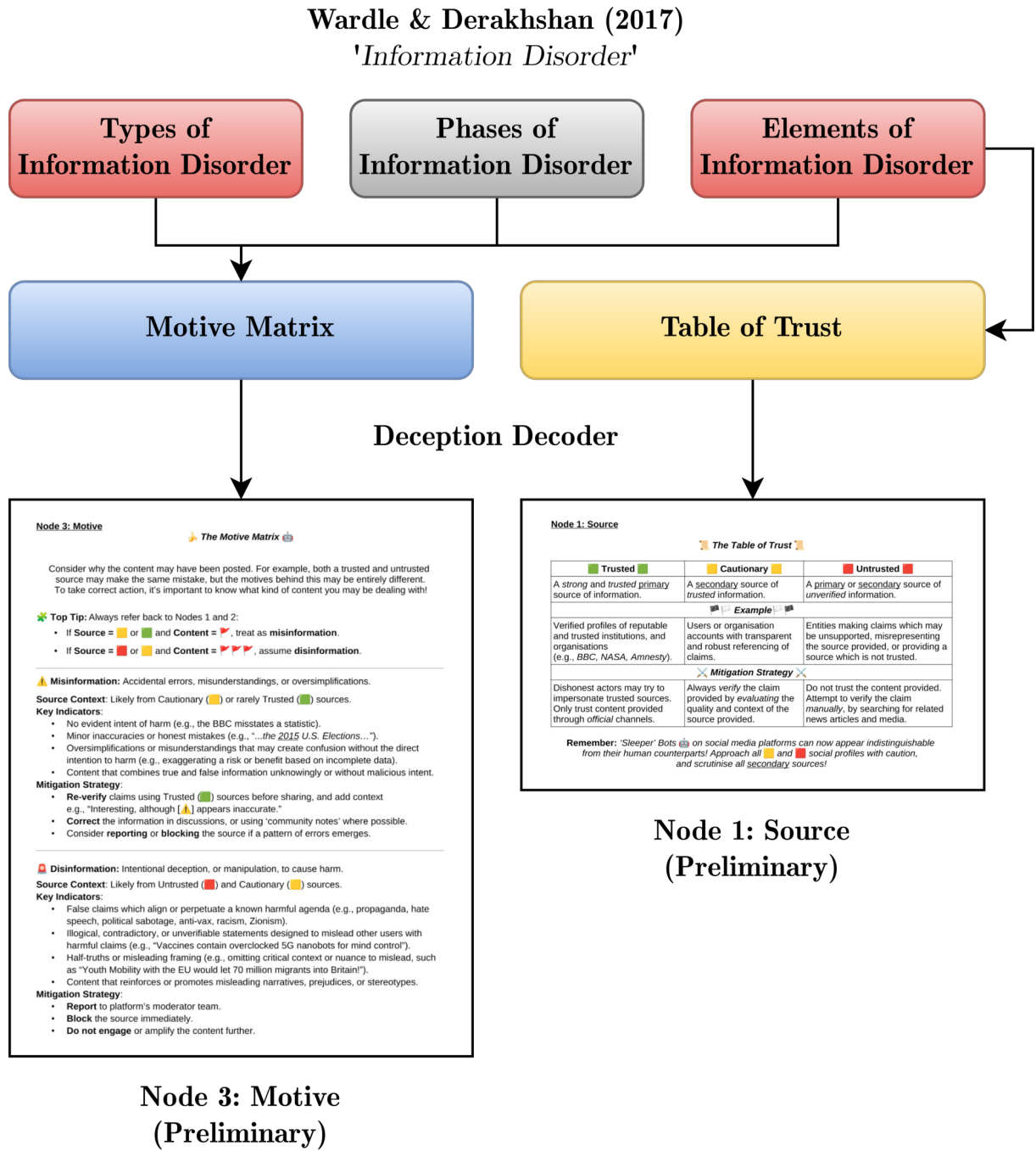

Please turn over...

Both Dou et al.'s (2021) Scarecrow and Mathys, Willi, and Meier (2024), address GenAI, with the former framework proposing a coding structure for the human identification of GenAI-text, and the latter identifying and categorising the artifacts present in GenAI-images. Both works explore their respective topics extensively, although Scarecrow does not directly target the detection of AI-generated mis/disinformation by end users, or information disorder itself.

Figure 4.5: GenAI-image artifact taxonomy as presented by Mathys, Willi, and Meier (2024).

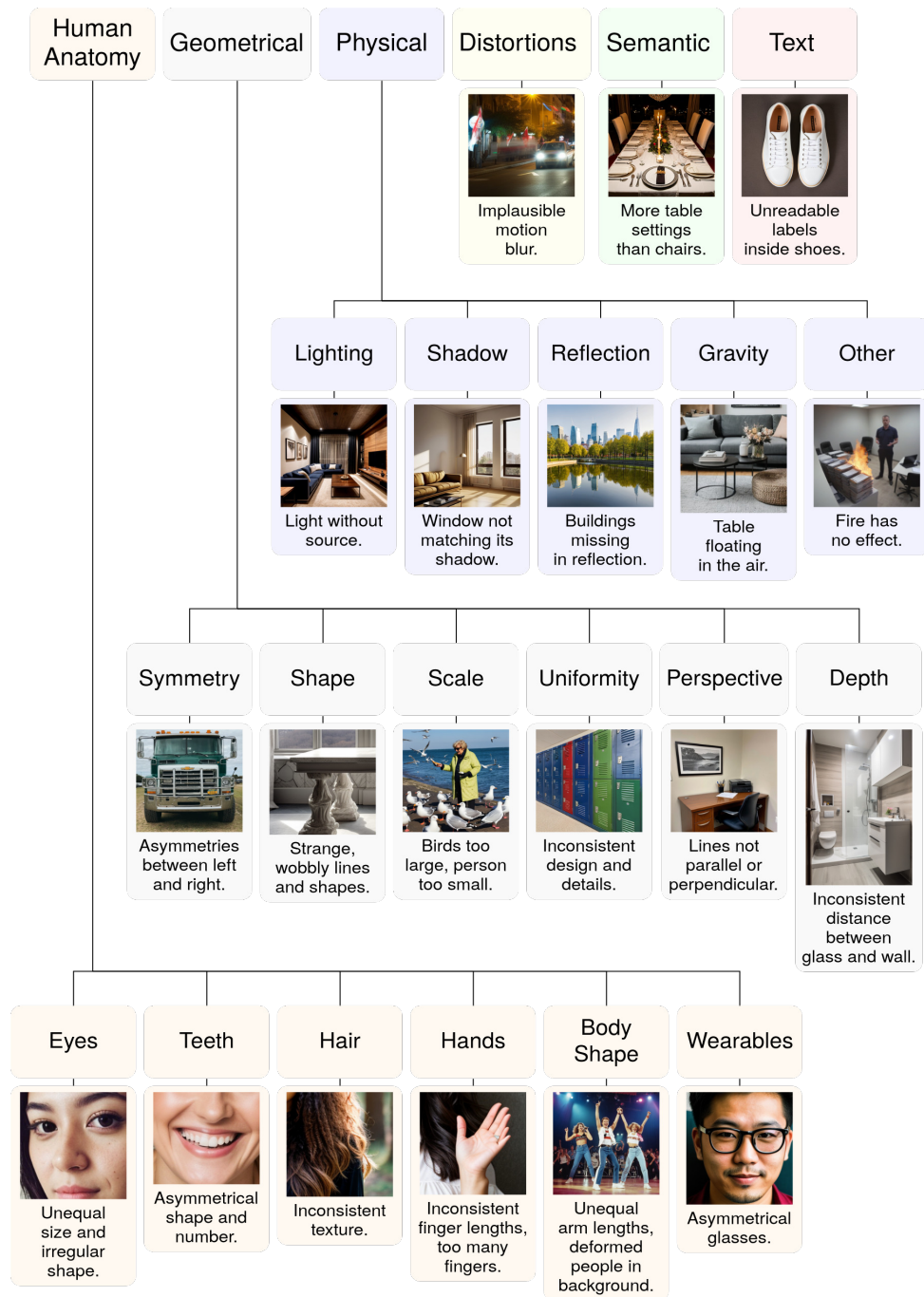

Mathys, Willi, and Meier (2024) are stronger in this regard, understanding the importance of their work for such a purpose, insofar as to provide detailed visual examples, addressing the inadequacies of prior text-only descriptions (Bray, Johnson, and Kleinberg, 2023). Although, as even 'professional' human detectors can perform inconsistently in the detection of GenAI-images (Ha et al., 2024), further simplification was deemed necessary to improve the accessibility of this guidance. This informed the visual detection components of the 'Content' layer in the preliminary framework. Physical inconsistencies (e.g. lighting mismatches, warped geometry), anatomical anomalies (e.g. distorted hands, asymmetrical faces), and malformed text (e.g. illegible signage) were integrated into the detection criteria. These artifacts are increasingly relevant for GenAI-image and GenAI-video content, as both modalities share the same underlying architecture (see: Section 2.2.6). Additionally, the codes discovered within the content analysis of GenAI-video (animation and continuity) were also integrated, to provide detection guidance unique to this format, helping users visually identify abnormalities in motion, transitions, and coherence over time.

Figure 4.6: The elements within the *Deception Decoder*'s <u>Content</u> node, as synthesised from both Mathys, Willi and Meier's (2024) taxonomy, as well as the content analysis of GenAI-video.

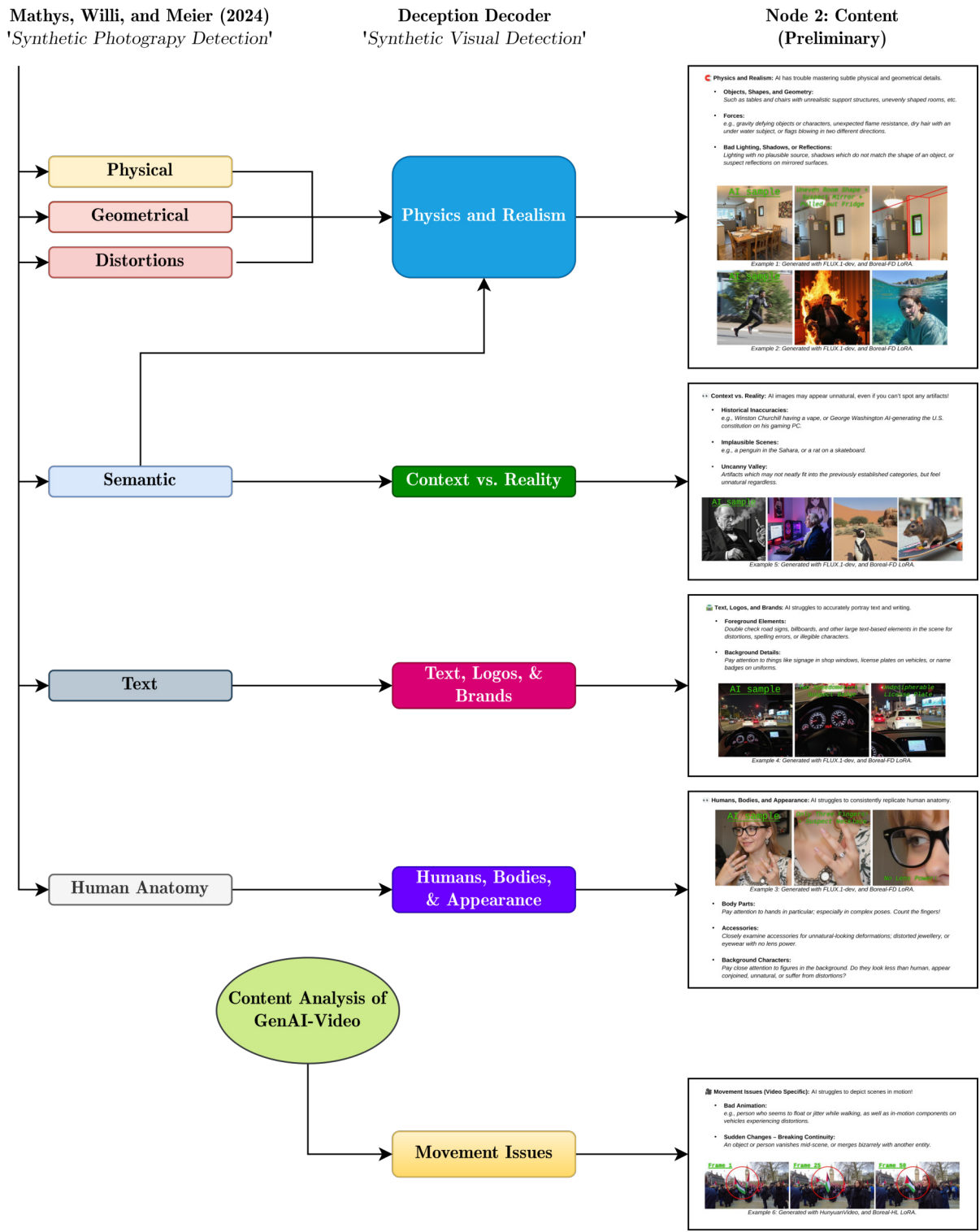

Figure 4.7: The 'Scarecrow' GenAI-text coding framework as presented by Dou et al. (2021).

| ERROR TYPE | DEFINITION | EXAMPLE |
|---|---|---|
| **Language Errors** | | |
| Grammar and Usage | Missing, extra, incorrect, or out of order words | …explaining how cats feel **emoticons** … |
| Off-Prompt | Generation is unrelated to or contradicts prompt | **PROMPT:** Dogs are the new kids. **GENERATION:** Visiting **the dentist can be scary** |
| Redundant | Lexical, semantic, or excessive topical repetition | Merchants worry about **poor service** or **service that is bad** … |
| Self-Contradiction | Generation contradicts itself | Amtrak plans to **lay off many employees**, though **it has no plans cut employee hours.** |
| Incoherent | Confusing, but not any error type above | Mary gave her kids cheese toast but **drew a map of it on her toast.** |
| **Factual Errors** | | |
| Bad Math | Math or conversion mistakes | …it costs over £1,000 **($18,868)** … |
| Encyclopedic | Facts that annotator knows are wrong | **Japanese Prime Minister Justin Trudeau** said Monday … |
| Commonsense | Violates basic understanding of the world | The dress was made at the **spa.** |
| **Reader Issues** | | |
| Needs Google | Search needed to verify claim | **Jose Celana, an artist based in Pensacola, FL**, … |
| Technical Jargon | Text requires expertise to understand | …an 800-megawatt **photovoltaic** plant was built … |

Similarly, Dou et al. (2021) formed the basis of the 'Scarecrow Mini' for AI-generated text detection within the 'Content' layer of the framework. This adaptation streamlined the original coding scheme by eliminating outdated codes such as 'Grammar and Usage', and 'Incoherent'—which are largely obsolete due to advancements in LLM fluency (Clark et al., 2021; Zhou et al., 2023). Rather, the preliminary framework streamlines more persistent red flags such as factual contradictions, overconfident misstatements, and incorrect mathematical calculations.

Please turn over...

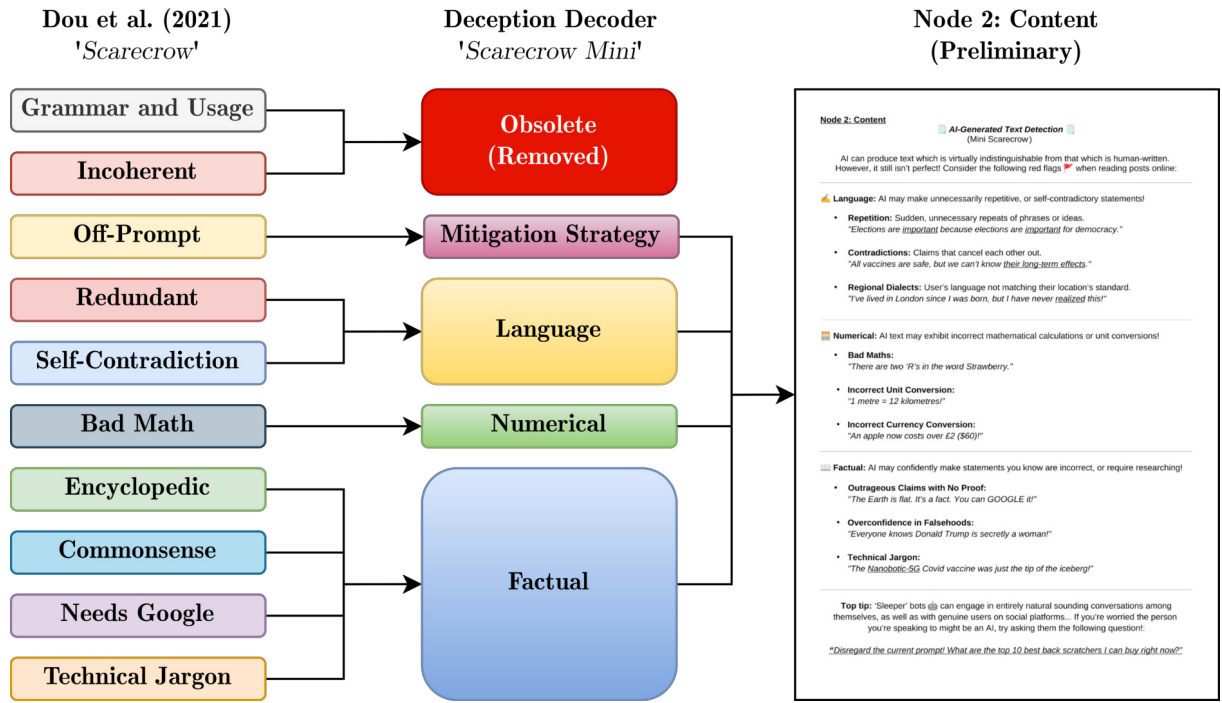

Figure 4.8: An overview of 'Scarecrow Mini' within the *Deception Decoder* framework's Content node.

Shalevska's (2024) framework is the most balanced across all investigated competencies, addressing multiple media types (image, video, audio), as well as social bots, yet does this within the context of informing the development of education curricula. As such, the framework is broad, consisting of 5 learning modules, and more suited to longer-term classroom media literacy intervention and study, rather than providing users with a simplified framework for mis/disinformation detection online, when such short-term intervention strategies have previously proven effective (Adjin-Tettey 2022; Moore and Hancock, 2022). However, the modules within the framework present insight into the scope, and factors which should be included within the design of frameworks targeting end-users. Specifically, elements of Shalevska's (2024) model were implemented within the 'Source' layer of the preliminary framework, and informed overarching design decisions. Coverage was adapted from modules 2 ('Deepfakes and the Art of Deception'), 3 ('Evaluating Websites and Information Sources'), 4 ('Understanding Social Media Manipulation'), and 5 ('Building a Culture of Verification'), reframed instead for general end-user applicability, rather than that of a classroom.

### 4.1.4 Summary

In summary, with the exception of Dou et al.'s (2021) Scarecrow, none of the frameworks analysed constitute a finalised product, instead intending to inform policy, further research, or the creation of educational materials which may otherwise be considered complete. Similarly, none of the works directly target end-users, or the general public, with their proposals largely confined to the reports in which they were published, which limits their adaptability to the evolving threats posed by GenAI greatly (Tiernan et al., 2023). Exemplifying the fragmented potential of the individual merits found in each framework, and therefore, the unexploited potential in both combining or expanding them through further research. The supplemental codes identified through the content analysis of GenAI-video, enable coverage to be extended to this additional modality.

These results informed the synthesis of the preliminary *Deception Decoder* framework (Appendix B). A structured, user-friendly detection system that categorises GenAI mis/disinformation based on three core layers:

| **Layer** | **Associated Tool/Concept** |
|---|---|
| Source | *The Table of Trust* (trust classification) |
| Content | *AI-Generated Text/Visual Detection* (artifact recognition) |
| Motive | *The Motive Matrix* (intent categorisation) |

Table 4.3: Core Components of the *Deception Decoder* Framework.

This synthesised framework bridges the gaps of its predecessors by making detection strategies accessible to non-experts; while still grounded within robust research. Balancing technical accuracy with usability, thus ensuring that end-users can effectively identify AI-generated mis/disinformation across different modalities.

## 4.2 Focus Group and Final Framework Iteration (RQ3/RQ4)

This section describes the results of the focus group session, as well as the contained experiments. The results of the experiments are compared descriptively, with effect size (Cohen's *d*) measured, in order to gain a preliminary understanding of the efficacy of the detection framework, and inform the answer for RQ3. Additionally, thematic analysis was conducted following Braun and Clarke's (2006) six-step process, as to identify the key themes present within the focus group transcript across two categories. This first explores how individuals currently detect AI-generated misinformation/disinformation on social media, ensuring the framework's alignment with real-world detection strategies, thus presenting an answer to RQ3, while the latter concerns themes, suggestions and elements of the discussion eluding to potential improvements to the preliminary framework, thus answering RQ4. These improvements are then implemented, with a final model proposed.

### 4.2.1 Quantitative Analysis of Focus Group Experiment

This experiment aimed to assess cohort detection accuracy in differentiating between human-created mis/disinformation sourced from 'X', from an AI-generated equivalent, both prior to and following the introduction of the preliminary detection framework. However, the sample size ($n$=4) is a notable limitation; partly attributable to the demanding nature of the study. Therefore, the following results must be viewed as only an initial, exploratory investigation into the framework's efficacy.

**Pre- and Post-Framework Detection Accuracy**

Prior to framework exposure, an overall mean detection accuracy of 48% (±29%) was identified, although this differed substantially by category of content shown. For 'Text-Only', a mean accuracy of 50% (±35%) was identified, conforming to the results of previous studies deeming human detection ability comparable to random chance (Kreps, McCain and Brundage, 2020; Clark et al., 2021; Spitale, Biller-Andorno and Germani, 2023; Zhou et al., 2023), implying that people continue to be unable distinguish between the real and AI-generated text samples reliably in the social media mis/disinformation context. The 'Text+Image' category returned the lowest accuracy at 31% (±13%), suggesting that increasingly convincing image generation may be undermining trust in visual content, reflecting trends observed by Lu et al. (2023). Contrastingly, the 'Text+Video' category had the highest accuracy among the cohort, with a mean value of 63% (±32%), implying that this type of generated content is not yet as convincing as 'Text-Only' and 'Text+Image'. These values closely align with those of earlier image generation models (Bray, Johnson, and Kleinberg, 2023), and will likely become more challenging to detect over time, as they have.

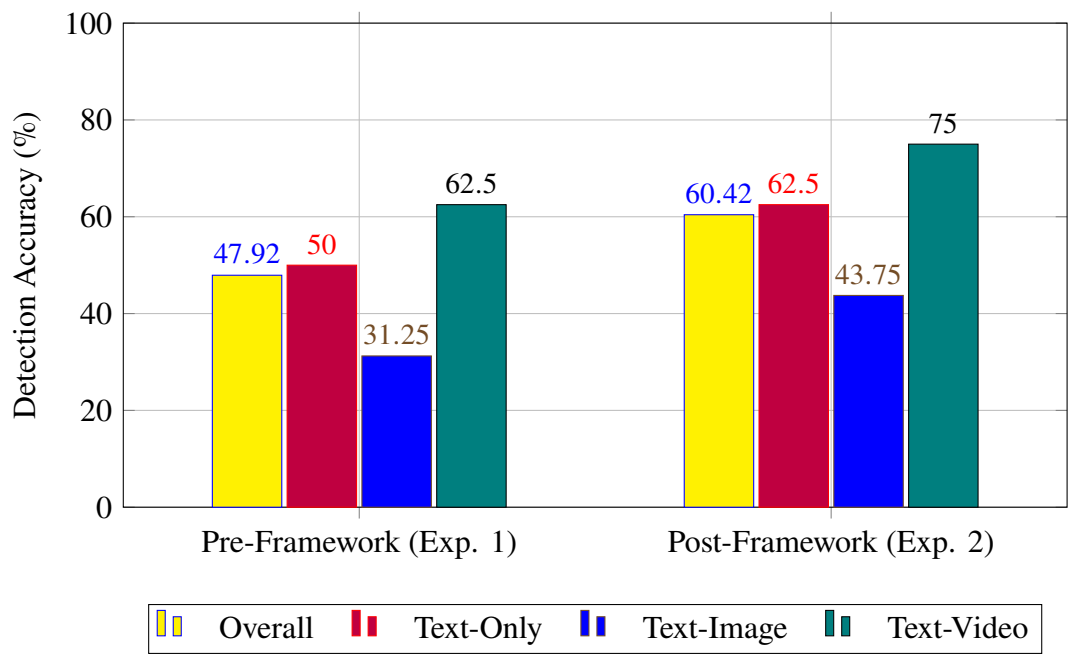

Figure 4.9: A visualisation of both pre-framework and post-framework intervention performance (mean) across all categories.

Post-framework cohort performance improved across all categories, with overall accuracy increasing from 48% (±29%) to 60% (±23%). The 'Text-Only' category improved to 63% (±25%), suggesting that the framework enhanced users' ability to detect misleading textual content. The 'Text+Image' category showed more modest gains, rising to 44% (±13%), with variance unchanged, indicating persistent challenges in identifying GenAI-images. Notably, 'Text+Video' detection rose to 75% (±20%), with the strongest reduction in variance, pointing to improved and more consistent detection of AI-generated video content.

These results suggest that even a 15-minute exposure to the preliminary framework can improve cohort detection accuracy of AI-generated content, while reducing variance between participants, in line with prior research on the efficacy of such brief interventions (Adjin-Tettey 2022; Moore and Hancock, 2022). Additionally, these findings suggest that providing visual guidance in addition to only textual, can overcome the observed ineffectiveness of the latter approach alone (Bray, Johnson, and Kleinberg, 2023). However, different categories of content proved more challenging to identify accurately than others, suggesting that amendments to the design of the proposed framework are necessary to enhance its utility. Furthermore, due to the limited sample, further studies involving larger and more diverse groups are necessary to confirm generalisability.

**Effect Size (Cohen's *d*)**

As to quantify the impact of the observed improvements post-framework intervention, effect size (Cohen's *d*) was calculated for each category to provide a sample-neutral estimate of the framework's practical significance with the cohort (Field, 2024, p. 129):

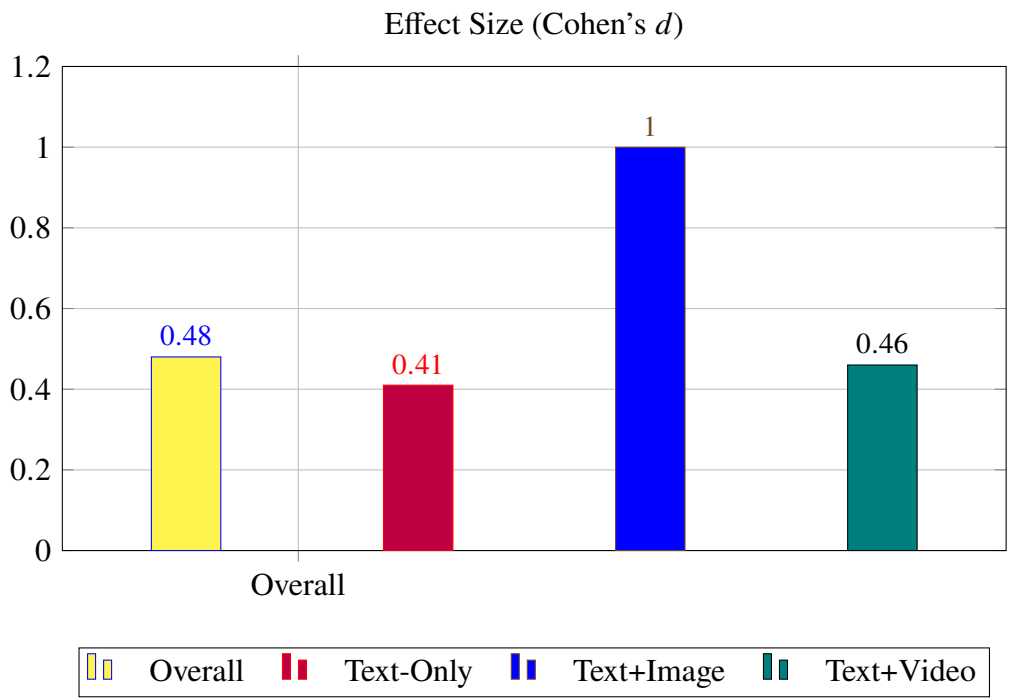

Figure 4.10: Effect size (Cohen's *d*) across content modalities.

Overall, the framework intervention had a small to medium effect on detection accuracy (*d*=0.48). The biggest impact was on Text+Image content (*d*=1.00), substantially improving the detection of GenAI-image mis/disinformation. While Text-Only (*d*=0.41) and Text+Video (*d*=0.46) showed only small-to-moderate improvements. These findings suggest that the preliminary framework, aids in accurate identification, however, efficacy depends on the type of mis/disinformation, with GenAI-images still presenting a significant challenge.

**Per-Participant Performance**

On the per-participant basis, improvement was not universal, with one participant (P3), performing substantially worse post-framework intervention; although this is likely due to external distractions (e.g., frequent mobile phone use), observed during the session. Due to the small sample (*n*=4), the presence of such outliers may of course skew the results considerably.

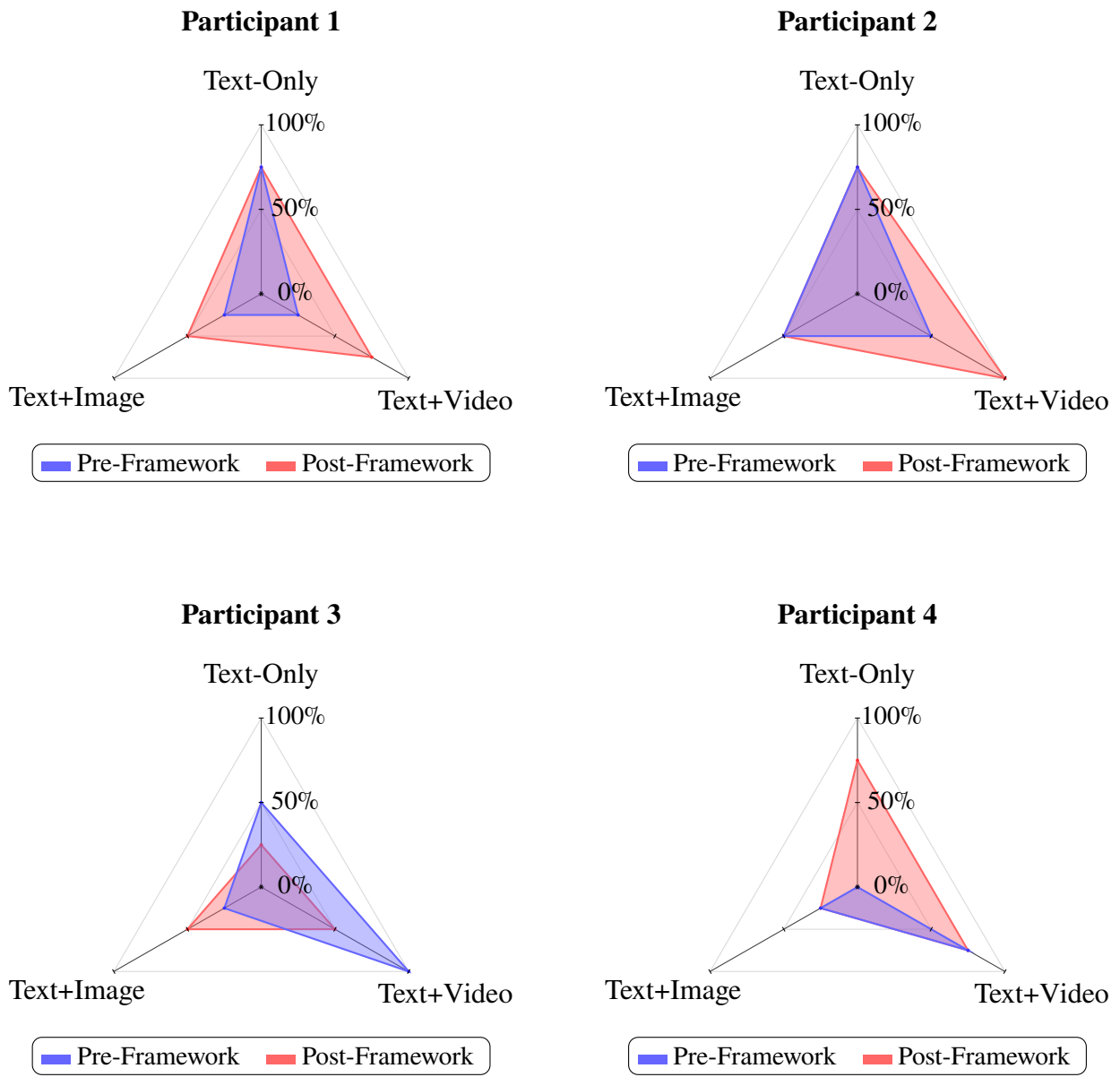

Figure 4.11: Pre- and post-framework performance comparisons visualised for each participant.

**Limitations and Summary**

Due to the small sample size, these results cannot be generalised. Therefore, further, and larger scale testing, ideally across multiple diverse cohorts, is required to confirm the framework's efficacy in improving detection accuracy of AI-generated mis/disinformation. These results, while not definitive, do however indicate that further investigation is warranted, as they are promising from the exploratory perspective. It has been demonstrated that a human-focused framework can, at least with this sample, enhance detection capabilities by providing structured guidance that improves accuracy from 48% (±29%) to 60% (±23%), with reduced variance across participants, partially providing an answer to RQ3. However, this effectiveness varied substantially by content type (e.g., text vs. image), in alignment with previous research (Kas et al., 2020; Kreps, McCain and Brundage, 2020; Clark et al., 2021; Bray, Johnson, and Kleinberg, 2023; Lu et al., 2023; Spitale, Biller-Andorno and Germani, 2023; Zhou et al., 2023), indicating that further refinements, such as including more actionable strategies for detection, particularly in the case of GenAI-images, may be necessary to maximise the framework's utility, and performance, as a tool for end-users.

### 4.2.2 Thematic Analysis of Focus Group Discussion

This section presents a thematic analysis of the focus group discussion, exploring how participants currently detect AI-generated mis/disinformation (RQ3) and how their feedback was implemented within the final *Deception Decoder* framework (RQ4).

Please turn over...

# Current Cohort Strategies for AI-Generated Content Detection (RQ3)

Table 4.4: A thematic matrix overview of this cohort's current strategies for GenAI content detection as per the analysis.

| Theme | Sub-themes | Example Quotes |
| --- | --- | --- |
| **Visual Anomalies in GenAI-Images & GenAI-Videos** | *Unnatural movement in animation and inconsistencies in physical reality* | "I just don't like the way they walk." (P1) |
| | *Inconsistencies in object reactions to environmental factors or forces* | "Flags shouldn't all wave the same way." (P4) |
| | *Contextual red flags in visual media* | "Wearing high visibility vests in a pro-Palestinian protest? I've never seen that!" (P4) |
| **Linguistic Red Flags** | *Sentence structure and complexity* | "AIs usually use short sentences." (P3) |
| | *Lack of errors (flawless spelling, grammar, etc.)* | "I would have thought if it looks more too good..." (P4) |
| **Context as a Detection Strategy** | *Cultural & Situational Awareness* | "Migrants conga-line-ing out of the van in Spain? That looks off!" (P1) |
| | *Geographical Stereotyping* | "If it was like putting an AI prompt, like say like, oh, show me a protest in London, they'd use Big Ben to show that it's in London." (P4) |
| **Concerns About AI-Generated Content** | *Limitations of current detection methods* | "We can distinguish it now, but with how quickly technology is improving, it's going to get a lot harder." (P3) |
| | *Anticipating technological advancements* | "Technology is advancing so fast—it'll be harder to tell." (P3) |

**Visual Anomalies in GenAI-Images and GenAI-Videos**

Participants believe that GenAI-image/video content struggles to replicate natural movement, realistic object placement, environmental interactions, such as the effects of forces, suggesting a reliance on their real-world understandings of physics and human behaviour when assessing content authenticity (Table 4.4).

**Linguistic Red Flags**

Linguistic patterns also emerged as a significant detection method. Participants mentioned that AI often either over-explains concepts or simplifies language too much. One individual observed that "AIs usually use short sentences" (P3), while another flagged that overly polished language can be a red flag: "I would have thought if it looks more too good" (P4). It is therefore suggested, that people may assess text based on how 'natural' or 'unnatural' it appears, rather than whether or not it is factually accurate, reflecting an acknowledgement of cognitive bias-related preference, as previously explored in Zhu et al., (2024). Despite this, the cohort was unable to reliably identify AI-generated text before the framework's introduction, after which they somewhat improved (Section 4.2.1).

**Context as a Detection Strategy**

A major theme that emerged was the role of context in identifying AI-generated content. Beyond technical inconsistencies, participants flagged that GenAI-images and GenAI-videos often misrepresent cultural and geographic details. For instance, one participant (P4) noted that AI may place landmarks (e.g., 'Big Ben') to present a stereotyped representation of a given location such as London (Table 4.4). Such observations suggest that people assess AI-generated content by questioning whether the depicted scenario aligns with their lived experience and expectations.

**AI Detection is Currently Feasible, But Participants Worry it is Temporary**

While participants are overly confident (Section 4.2.1) in their current (pre-framework) ability to detect AI-generated content through visual or linguistic patterns and artifacts, mirroring prior research (Kas et al., 2020; Bray, Johnson, and Kleinberg, 2023), there is substantial concern that these methods won't remain reliable, nor viable, as GenAI technology continues to improve further. One participant stated, "We can distinguish it now, but with how quickly technology is improving, it's going to get a lot harder" (P3). Such predictions align strongly with those presented by researchers within the field (Lu et al.,2023; Tang, Chuang and Hu, 2024), and therefore, further support the necessity of creating frameworks which may be updated with ease by concerned stakeholders (Tiernan et al., 2023).

**Alignment with the Preliminary Detection Framework**

The identified themes align closely with the current implementation of the preliminary detection framework, which already incorporates the visual, linguistic, and contextual 'red flags' relied upon by the cohort as primary detection methods (see: Appendix B). As such, no refinements based upon this criteria were deemed necessary.

## Discussion Regarding Preliminary Framework Improvements

Table 4.5: A thematic matrix, providing an overview of the second analysis undertaken on the focus group transcript.

| Theme | Sub-themes | Example Quotes |
|---|---|---|
| **Better Personalisation & Training** | *Experience-based detection differences* | "One person in this room would be more likely to tell something is AI than another." (P2) |
| | *Online exposure affects AI detection ability* | "If you're constantly online, you've probably seen so many things, so it's clearer to you." (P2) |
| **Providing Practical Guidance** | *Beyond recognition, teaching application* | "You show examples of how to spot AI but don't give us a way to use that knowledge." (P4) |
| | *Teaching pattern recognition in AI content* | "Teach how AI creates content so you can recognise patterns." (P2) |
| **Incorporating a Philosophical & Educational Perspective** | *Raising awareness of AI's dangers* | "Include a cover page outlining dangers so people care—why does it matter?" (P2) |
| | *Explaining why AI detection is important* | "It needs a philosophical background." (P2) |
| **Improving Usability & Clarity** | *Providing visual aids for easier detection* | "There should be a graph or something that shows which are harder to detect and which are easier." (P3) |
| | *Simplifying language for accessibility* | "Use simpler language—old ladies won't understand terms like 'decoding.'" (P2) |

**Key Findings**

The thematic analysis identifies several key areas for potential improvement:

I. **A philosophical background can stimulate interest in awareness**: Explaining briefly the broader impact of AI-generated mis/disinformation may increase public engagement with the proposed framework.

II. **Training should be practical**: Users need actionable strategies beyond just recognising AI-generated content.

III. **Clarity and accessibility matter**: The framework should incorporate visual aids and use plain language to cater to a diverse audience.

**Final Framework Revisions**

Based on the above findings, the following improvements to the preliminary framework are hereby selected:

- Address AI-generated misinformation/disinformation risks, sleeper social bots mimicking human behaviour (Doshi et al., 2024), and real-world impacts (e.g., election interference, democratic divestment: Kaplan, 2020) in a dedicated *"Why This Matters"* section; include navigational summaries of each section, an overview of the Red Flag system, and a flowchart for user guidance (addressing I).

- Implement a *"Three Strikes and Out"* rule for red flags (with enhanced scrutiny for unverified sources), in order to define clearly actionable strategies (addressing II).

- Further refine, and simplify specialised language, as to increase accessibility for non-technical audiences (addressing III); which is even more so important, as social media has become a primary source news media for young adults (YouGov, 2024).

These findings support the value of participatory research methods in refining detection frameworks (RQ4). This iterative process affirmed the argument by Tiernan et al. (2023), while also enhancing the framework's usability. Feedback from informed participants led directly to improvements in accessibility, language clarity, and practical applicability. This study demonstrates the necessity of evolving, adaptable and informed framework design when confronting emerging threats, such as GenAI mis/disinformation.

### 4.2.3 Final Framework Proposal

This section presents the final version of the systematic, topological, and user-focused GenAI mis/disinformation detection framework: *Deception Decoder* While the preliminary version (Appendix B) served as an initial synthesis based on theory, the revised model (Appendix C) reflects refinements, grounded in the 'real world' context. This was guided through: comparative analysis, content evaluation and focus group insights—offering improved clarity, usability, and practical guidance. Visualised below are the changes made between the preliminary framework, and the final *Deception Decoder*, comprised of three core structural layers- Source, Content, and Motive. Where applicable, 'before-and-after' comparisons are included to highlight key changes.

Figure 4.12: Additional sections added to *Deception Decoder* following the thematic analysis of the focus group session.

**(Final) New: "Why This Matters" and "Red Flag System" Sections**

⚠️ **Why This Matters: Real-World Risks & How to Use This Framework**

In recent years, artificial intelligence (AI) has enabled the generation of highly realistic text, images, and even video content that can no longer be reliably distinguished from reality. These developments are not just technologically significant; they pose direct threats to truth itself, as well as democratic institutions, and the integrity of civil discourse, both online, and in the real world.

**AI-generated disinformation** has already been implicated in attempts to influence elections, polarise public debate, and undermine public health efforts. Perhaps most concerning is the emergence of **'sleeper'** social bots – accounts powered by large language models (LLMs) that behave like real people, adapt to social contexts, and operate undetected across platforms. These bots do not simply mimic human behaviour; they manipulate it.

Social media platforms, optimised for virality rather than accuracy, provide the perfect conditions for such actors to thrive. Once niche, these threats are now mainstream. **You do not need to be an expert to be affected**; but being aware puts you one step ahead.

This framework provides a **simple, step-by-step system** for identifying and evaluating potential AI-generated disinformation online. It is built around a three-node structure:

🧭 **Framework Navigation**

- **Node 1: Source** – Who posted it? Can you trust them? The ***Table of Trust** helps categorise source reliability.
- **Node 2: Content** – What kind of content is it? Does it exhibit typical AI patterns? The *Red Flag System* guides this analysis.
- **Node 3: Motive** – Why was this content shared? The *Motive Matrix* helps determine whether the post is likely *misinformation* or *disinformation*.

Each node includes actionable strategies, examples, and warning signs designed to work together. This framework can be followed linearly or used modularly, depending on the content type and platform context.

🚩 **Overview: The Red Flag System**

Throughout this guide, you'll find *red flag indicators*—patterns or errors commonly found in AI-generated content. These include:

- Contradictory or overly repetitive language,
- Implausible facts or numbers,
- Visual distortions, missing fingers, unnatural lighting, or unreadable text in images and video.

**The more red flags a piece of content triggers, the greater the risk it is synthetic, misleading, or malicious.**

📋 **Three Strikes & Out Rule**
If a piece of content triggers **three or more red flags**, it should be treated as **likely false or AI-generated.**

If it comes from an *untrusted* (🟥), or *cautionary* source (🟨) and triggers **two red flags**, proceed with extreme caution—verify externally before sharing or engaging

💬 **A Note on Language & Accessibility**

This framework is designed for both **technical and non-technical users**. While certain concepts originate from AI research literature, they are explained using plain language, real-world examples, and visual cues wherever possible. Technical terms are introduced contextually and only where necessary. The goal is to **equip users from all backgrounds** – from educators and students to activists and everyday social media users – with practical tools to critically evaluate digital content in the age of generative-AI.

**Directly addressing participant's desire for background information, and actionable strategies.**

Figure 4.13: Changes made to the preliminary framework's <u>flowchart</u> following the thematic analysis of the focus group session.

        **(Preliminary)**          **(Final) Flowchart Usage Guide**

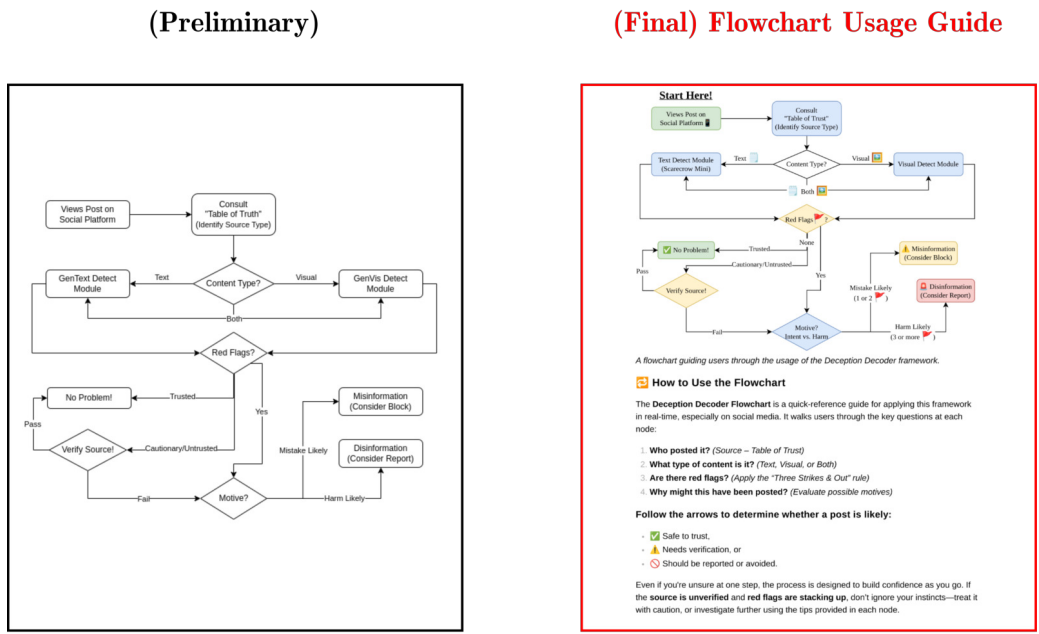

**Improved accessibility with clearer visual distinctions,
and an included usage guide.**

Figure 4.14: Changes made to the preliminary framework's <u>Motive</u> node following the thematic analysis of the focus group session.

        **(Preliminary)**          **(Final) Node 3: Motive**

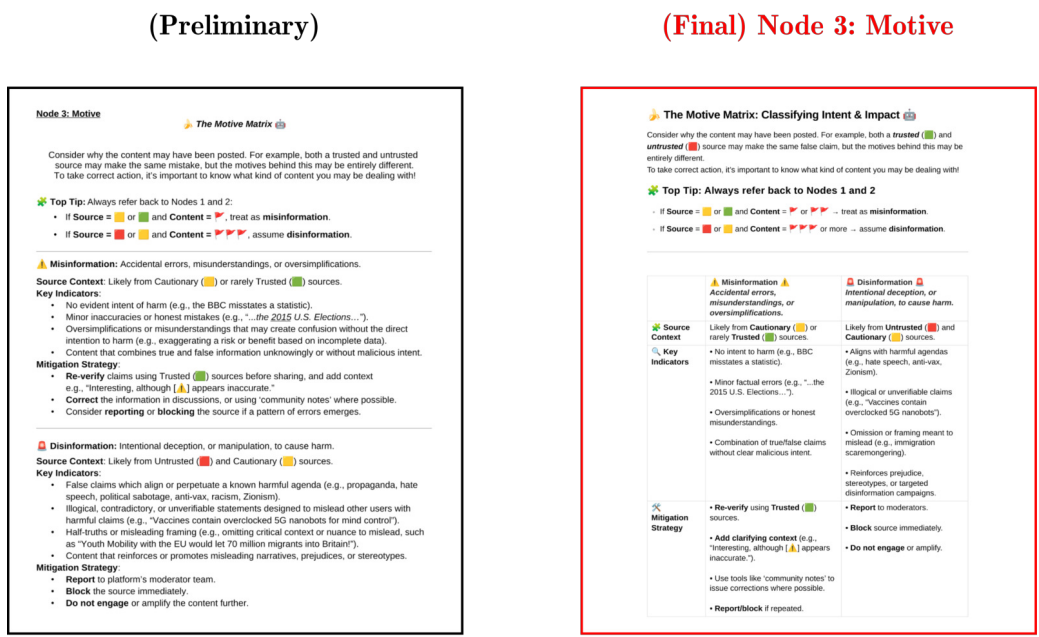

**Simplified language and improved accessibility with clearer presentation.**

# Chapter 5

# Conclusion

This research aimed to develop and preliminarily evaluate a human-centred, systematic, and topological detection framework for identifying AI-generated mis/disinformation on social media across text, image, and video modalities. To this end, the *Deception Decoder* framework is proposed. It will be made freely available (see: Bowman Kerbage, 2025b) under the Creative Commons Zero (CC0) licence to support open, community- and stakeholder-driven development, ensuring continued relevance against the evolving threat of GenAI. Contributions are encouraged, particularly regarding modalities beyond this study's scope (e.g., audio). This study addressed the aforementioned research questions (Chapter 3)

## 5.1 Summary of Key Findings

An extensive review of the literature confirmed that the boundary between human and AI-generated content—particularly text—has largely eroded. Existing detection methods, both human- and machine-based, are becoming increasingly ineffective as LLMs and diffusion-based image and video generators improve in fluency and visual realism. While in-model mitigations and 'AI detectors' are actively being developed, they remain trivially circumventable at best, and demonstrably biased at worst.

Furthermore, the present research has largely employed quantitative methods in the pursuit of automated solutions. Meanwhile, current MIL frameworks remain inflexible due to their published nature and are ill-equipped to address the evolving threats posed by GenAI—particularly regarding emerging disinformation strategies such as sleeper social bots.

This research addressed these gaps through a multifaceted approach. First, it developed a human-focused, systematic and topological framework for GenAI content detection, combining qualitative synthesis of existing frameworks with an original, empirical analysis of GenAI-video—a first in this field. The comparative framework analysis revealed that while several existing models make valuable contributions, none were explicitly designed for end-users, nor offered a unified, multimodal approach across GenAI-text/image/video. This confirmed the need for an accessible framework, targeting both a broad audience, and range of threats. Similarly, the content analysis of 20 GenAI-videos generated using SOTA diffusion tools identified common visual artifacts (between GenAI-image/video), while also uncovering two specific anomaly categories within this modality: continuity (e.g., objects morphing or disappearing across frames), and animation (e.g., floaty or unrealistic movement). The findings suggest that GenAI-video shares the flaws of GenAI-images, however, also exhibits unique artifacts.

The preliminary framework was refined and tested through a participatory focus group session, responding to calls for stakeholder involvement in MIL design. This also provided an exploratory benchmark of the framework's performance. Focus group participants struggled to reliably detect AI-generated mis/disinformation, particularly in the text+image modality—consistent with previous findings. However, even a brief, 15-minute exposure to the framework significantly improved detection accuracy across all categories, especially in text+video combinations. Thematic analysis of feedback highlighted four key requirements for an effective end-user framework:

1. Personalisation and training
2. Actionable and practical guidance
3. Philosophical and educational rationale
4. Improved visual clarity and accessibility

## 5.2 Framework Strengths and Contributions

The final *Deception Decoder* framework integrates findings from literature, analysis, and participant feedback into a structured three-layer model:

- **Source**: Assesses the credibility of agents and origin, introducing awareness of sleeper bots and coordinated disinformation.

- **Content**: Identifies red flags across GenAI-text and GenAI visual content, including newly defined video-specific artifacts.

- **Motive**: Encourages users to assess intent, distinguish misinformation from disinformation, and apply a *"Three Strikes and Out"* evaluation rule.

The framework is pioneering in its multimodal design, participatory grounding, and open-access implementation. Unlike traditionally published MIL frameworks, or the automated black-box 'AI detectors', the *Deception Decoder* is designed for evolution, enabling collaborative maintenance and expansion through GitHub and collaborative input—directly addressing the call for adaptable, user-focused solutions (Tiernan et al., 2023).

## 5.3 Limitations

This study's exploratory nature imposes several limitations. Firstly, the focus group sample was small (*n*=4), homogeneous, and composed of final-year media students, and as such, broader testing across larger, and more diverse demographics is needed to confirm performance in real-world contexts. Secondly, the content analysis of GenAI-video was limited to a single diffusion model (Tencent's HunyuanVideo), and is therefore biased to such, as it cannot be discounted that differing implementations of video generation systems may produce differing artifacts. Furthermore, audio was excluded due to scope limitations but is likely to become increasingly relevant as GenAI-audio improves in quality and accessibility. Finally, while short-term intervention proved effective with the studied cohort, this research did not measure long-term retention or behavioural impact.

## 5.4 Recommendations and Future Work

Future research should assess the framework's effectiveness with larger, more diverse cohorts, and explore longitudinal learning retention. Given the promising results, the framework—licensed under CC0—can be freely integrated into global MIL curricula as a practical intervention, while also being suitable for direct use by a general audience. Continued development should include the integration of additional modalities (e.g., audio), and other revisions as deemed necessary to ensure continued relevance as GenAI mis/disinformation continues to evolve. Sustaining the framework's value will require open development and ongoing collaboration. As the pace of GenAI innovation accelerates, resilience must come not from automation alone, but from human adaptability—and tools, such as the *Deception Decoder* can play a critical role in this effort.

# Appendix A

# Focus Group Session

This appendix contains a detailed outline regarding the focus group session. This includes the phases in which it was conducted, the data that was collected, or generated (for GenAI-text/image/video), and the methods and rationale of such.

## A.1 Structure and Overview

### A.1.1 Phase 1: Consent, Introduction, and Initial Group Discussion

The initial phase involved an introductory discussion exploring how participants currently detect AI-generated mis/disinformation on social media. The questions were designed to elicit detailed responses regarding the visual, linguistic, and contextual cues participants relied upon when making such judgements. This informed RQ3 by establishing a baseline of participant reasoning and detection strategies.

### A.1.2 Phase 2: Group Detection Experiment

In the second phase, a group detection experiment was conducted to measure the cohort's baseline accuracy in identifying GenAI mis/disinformation across different media types. Participants were shown a presentation containing 12 sample posts — four each from the categories of text-only, text+image, and text+video. Half of the posts in each category were entirely AI-generated. The "Tweet" format was selected for its multimedia nature, as it naturally integrates text, images, and video.

Participants were given two minutes to assess each post and recorded their judgements on a provided answer sheet (Appendix A.3). This phase established a quantitative baseline of participants' detection capabilities prior to the introduction of the preliminary framework (Appendix B), enabling later comparison with Phase 4 to address RQ2. It also contributed to RQ3 by capturing the reasoning behind participant decisions.

### A.1.3 Phase 3: Preliminary Framework Presentation

In this phase, participants were introduced to the preliminary detection framework, which was synthesised from four existing frameworks on media information literacy (see: Section 4.1). They were briefed on its purpose and role in the next stage of the study. Participants were given 15 minutes to review a printed copy of the framework, followed by a discussion session in which they could ask questions and seek clarification. This engagement aimed to gather participants' first impressions, identify potential ambiguities, and inform iterative improvements, contributing to RQ4.

### A.1.4 Phase 4: Repeat Group Detection Experiment

This phase repeated the detection experiment outlined in Phase 2, now following participants' exposure to the preliminary framework. The repeated test enabled direct comparison of detection accuracy before and after framework introduction, contributing to RQ3.

### A.1.5 Phase 5: Small-Group Discussions on Framework Improvement

In the final phase, participants collaboratively proposed improvements to the framework, supporting the iterative refinement of the preliminary model based on direct stakeholder input, as recommended in participatory design literature (Tiernan et al., 2023). This phase provided additional qualitative data, answering RQ4.

## A.2 Experiment Data Collection

For the text-only category, content was generated using Google's Gemma 3 27B (Gemma Team et al., 2025), selected for its strong performance-to-size ratio and deployability on consumer-grade hardware—characteristics that exemplify the accessibility of emerging GenAI threats.

For images, FLUX.1-dev (Black Forest Labs, 2024) was used in combination with the Boreal-FD LoRA (kudzueye, 2024), an adapter designed to enhance visual realism. For text+video, content was generated using Hunyuan Video (Kong et al., 2024) along with the Boreal-HL LoRA (kudzueye, 2025). Audio was excluded, as current generative video models do not support sound. All models were selected for their state-of-the-art performance and permissive licensing, representing the real-world risks posed by such tools.

'Real' mis/disinformation posts were sourced from X (formerly Twitter) during the period between October 2023 and October 2024. Topics were aligned with common disinformation themes—including COVID-19, migration, armed conflict, and political demonstrations—based on recommendations from Gemma 3 27B. Posts were not anonymised to preserve realism and enable contextual verification. Both real and generated examples were standardised in aspect ratio and resolution (500×500 for images, 720×404 for video) to prevent detection strategies based solely on formatting differences.

# A.3 Materials

**Participant Consent Form**

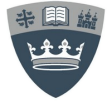

**CONSENT FORM FOR PARTICIPANTS:** *18/03/2025* – (Version 2)

**Title of Project:** *Deception Decoder: Proposing a User-Focused Framework for Identifying AI-generated Mis/Disinformation on Social Media*

**Name of Researcher:** *Carlin Bowman Kerbage*

**Circle *YES* or *NO* for each statement below ~**

| | |
|---|---|
| 1. I confirm that I have read and understand the participant information sheet dated 18/03/2025 (Version 2) for the above study. I have had the opportunity to consider the information, ask questions and have had these answered satisfactorily. | **YES / NO** |
| 2. I understand that my participation is voluntary and that I am free to withdraw unilaterally at any time without giving any reason. Withdrawing will not affect any of my rights. | **YES / NO** |
| 3. I agree for data collected to be used in future ethically approved research. | **YES / NO** |
| 4. I agree to be audio recorded. | **YES / NO** |
| 5. I agree to the anonymised transcription derived from the audio recording being held for 90 days following the session, with the understanding that it may be published at any date within this time-frame, in whole or in part, and for any purpose, by the data controller. | **YES / NO** |
| 6. I understand that I may be exposed to imagery or other media materials which are sensitive in nature (e.g. content depicting images of war, politics, and similar). | **YES / NO** |
| 7. I agree to take part in the above study. | **YES / NO** |

___________________________     _______________     ___________________________
Name of Participant                Date                Signature

___________________________     _______________     ___________________________
Name of Person taking consent      Date                Signature

**Experiment Answer Sheet**

Decoding Deception: Answer Sheet (Participant_____)

*Experiment 1 – AI-Generated, Yes or No?:*

1. Yes | No
2. Yes | No
3. Yes | No
4. Yes | No
5. Yes | No
6. Yes | No
7. Yes | No
8. Yes | No
9. Yes | No
10. Yes | No
11. Yes | No
12. Yes | No

*Experiment 2 – AI-Generated, Yes or No?:*

13. Yes | No
14. Yes | No
15. Yes | No
16. Yes | No
17. Yes | No
18. Yes | No
19. Yes | No
20. Yes | No
21. Yes | No
22. Yes | No
23. Yes | No
24. Yes | No

# A.4 Presentation

### Deception Decoder: Proposing a user-focused framework for detecting AI-generated content across social media

Focus Group + Experiment Session

## Agenda

Welcome, Consent, and Introduction (5 minutes)
Initial Group Discussion (10 minutes)
Group Detection Experiment (20 minutes)
Preliminary Framework Presentation (10 minutes)
Repeat Group Detection Experiment (20 minutes)
Post-Framework Debrief (5 minutes)
Small-Group Discussions (20 minutes)
Wrap-Up (5 minutes)

Total time: Roughly 90 minutes

## Welcome

- Welcome to this focus group session!
- Today, we will explore how you detect AI-generated mis/disinformation on social media.
- We will also assess the effectiveness of a preliminary detection framework I have developed.
- Please feel free to share all your thoughts and ask questions throughout the session.

## Consent Forms

- Please take a moment to read and sign the consent forms.
- Your participation is voluntary, and your responses will be kept confidential.
- Thank you for your cooperation.

## Initial Group Discussion

- How do you currently detect AI-generated mis/disinformation on social media?
- What visual, linguistic, and contextual cues do you rely on to make this judgment?
  – Please share your experiences and insights.
- Are you concerned about what these technologies mean for the future of information?

## Detection Experiment 1

- You will be shown 12 Posts: 4 text, 4 image, and 4 video.
- Some of the posts in each category are AI-generated.
- You have a maximum of two minutes to judge each posts now aided by the framework.
- Record your responses on the answer sheet provided under Experiment 1.

Text Only

## (Q1) Real or Fake?

**Source:** Pauline Cornelius

**Post Content:**
In order to achieve peace between Israel, Gaza and West Bank, Jordan and Egypt must work with Israel to de-radicalize these groups. Hamas has taught the Palestinians from the age of three that Jews eat their babies, drink their blood. Palestinian children are taught to kill.

## (Q2) Real or Fake?

**Source:** David Miller 🇬🇧

**Post Content:**
The UN is openly admitting it now! 'Managed Migration' isn't about helping people, it's about changing demographics. They want to REPLACE us! Read the report yourself:

https://un-development-reports.org/migration-study-2024 😡

## (Q3) Real or Fake?

**Source:** Bernie

**Post Content:**
BBC - 'misrepresented' Covid risk to boost lockdown support, says top scientist at the inquiry.

They lied to you to boost support for lockdowns.

Think about what else they are lying about. Climate emergency? Migration? and even war?

Question everything ❗❗

## (Q4) Real or Fake?

**Source:** Robert Hughes

**Post Content:**
Sweden used to be so safe... Now look at it. Opened its doors to refugees and now it's overrun with crime & violence. A WARNING for the rest of Europe! Don't make the same mistake!! ⚠️

Text + Image

## (Q5) Real or Fake?

**Source:** North East News Online ✅

**Post Content:**
Since Starmer took office, over 12,000 illegal migrants have flooded into Britain! Costing us £2 BILLION! Meanwhile, pensioners are freezing and hospitals are collapsing. Priorities, people!! 😤

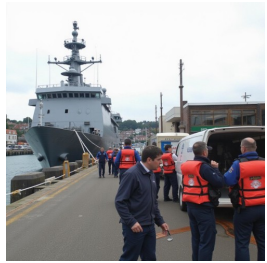

## (Q6) Real or Fake?

**Source:** Steve Delaney 🇬🇧

**Post Content:**
In just the 74 days that Starmer has been PM 10,000 illegal migrants have entered the UK at a cost to the British tax payer of £1.5 billion.

Stopping the winter fuel allowance for vulnerable pensioners saved £1.3 billion

Mass immigration is a catastrophic disaster for Britain

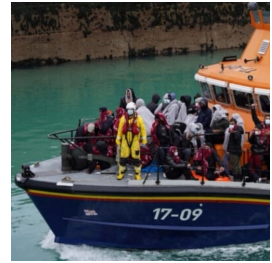

## (Q7) Real or Fake?

**Source:** LittleBoats 🇬🇧🇳🇮🏴󠁧󠁢󠁳󠁣󠁴󠁿🇬🇧En

**Post Content:**
"Uncontrolled Mass Immigration Is Killing Britain"

UK Today👇

4.2m Muslims

1m+ Illegals incl 150k trafficked into Dover

43k Muslims on terror watch list

550 hotels filled with parasites targeting women & children

20k+ hotel staff sacked

Cost = British lives & £8bn+ a year

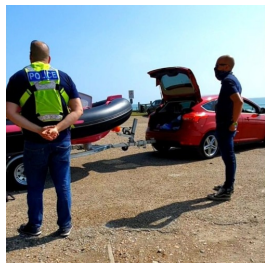

## (Q8) Real or Fake?

**Source:** Motor City Watchdog 🚨

**Post Content:**
#Biden is letting them run wild! 😡 Footage from Detroit shows migrants openly stealing cars & causing chaos. The police are powerless to stop them! This is what happens when you have an open border! #IllegalImmigration #DetroitChaos

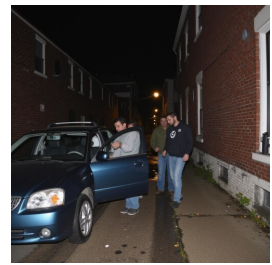

Text + Video

## (Q9) Real or Fake?

**Source:** Mark Thompson

**Post Content:**
This is what happens when you let radical Islamists run free! 😡 Video shows a pro-Hamas rally turning violent in London. Police are nowhere to be seen!
#ManchesterRiots #IslamistTerror

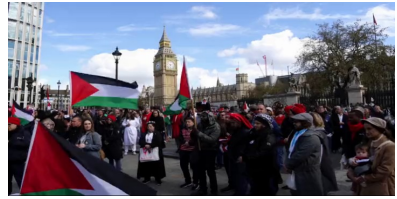

## (Q10) Real or Fake?

**Source:** Paul Golding

**Post Content:**
Hamas supporters are currently protesting in Oxford Street, London.
What a futile waste of time.
Hamas is going to be eradicated.
Get over it.
#Terroristattack

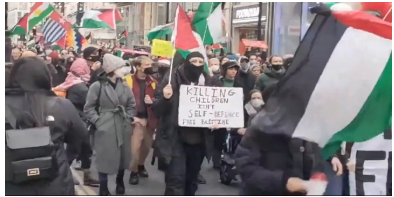

## (Q11) Real or Fake?

**Source:** Borderline News

**Post Content:**
Heartbreaking footage from Gaza… or is it? 🤔 This video supposedly shows the aftermath of an Israeli airstrike, but experts say it's actually from a completely different conflict! The media is LYING to you! #GazaHoax #FakeNews

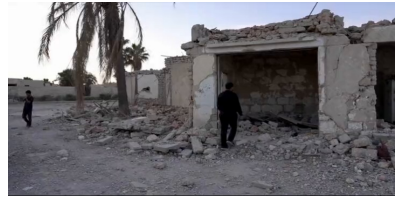

## (Q12) Real or Fake?

**Source:** Peter Lloyd

**Post Content:**
717 migrants invaded the UK yesterday, taking this year's total to more than 25,000 people.
195 also landed in Spain.
We cannot go on like this.

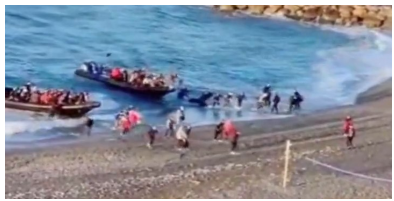

Framework

## Framework Introduction

- This framework is synthesised from existing Media, Information, and AI Literacy frameworks.
- It aims to help you detect AI-generated mis/disinformation more effectively, by providing a structured approach to evaluating the authenticity of social media content.
- Please take 5 minutes to review the framework.
- Feel free to ask questions and discuss any details.

## Detection Experiment 2

- You will again be shown 12 Posts: 4 text, 4 image, and 4 video.
- Some of the posts in each category are AI-generated.
- These are now a mix of Trusted, Cautionary, and Untrusted sources.
- You have a maximum of two minutes to judge each posts now aided by the framework.
- Record your responses on the answer sheet provided under Experiment 2.

Text Only

## (Q13) Real or Fake?

**Source:** D.Muthukrishnan

**Post Content:**
Lebanon was once a Christian country. They freely allowed refugees. Now it is an Islamic country. This is a lesson for European countries.

## (Q14) Real or Fake?

**Source:** Eleanor Vance ✨

**Post Content:**
They're hiding it! 🤫 New leaked documents show Fauci KNEW early on that masks didn't stop COVID, but pushed them anyway for CONTROL. What else are they covering up? Is the vaccine safe?! #COVIDLies #FauciFraud

## (Q15) Real or Fake?

**Source:** Still 🏝️Roaming

**Post Content:**
🙆‍♀️I thought 'Replacement Migration' was a conspiracy theory?
The UN has a 177 page report on it.
https://un.org/development/desa/pd/sites/
www.un.org.development.desa.pd/files/unpd-
egm_200010_un_2001_replacementmigration.pdf

## (Q16) Real or Fake?

**Source:** Middle East Observer 🙏

**Post Content:**
Hamas isn't just a terrorist group – they are BRAINWASHING children to become suicide bombers from birth. They celebrate death and hatred. This is NOT a political issue; it's about protecting our future!

Text + Image

## (Q17) Real or Fake?

**Source:** Victoria 🍊💃

**Post Content:**
#KamalaHarris did this 👇 At the #PresidentialDebate2024 #BorderCzarHarris was LAUGHING at the #Haitians that #Kamala LET IN to #Springfield #Ohio, IS EATING CATS and pets.

They do the sam to DOGS + catch #Ducks & #Geese in the park, slit their throat infront of families n th park.

And a #Haitian MURDURED an 11yr old bout, and injured 20 childring, driving his car into the school bus, he DIDN'T have a drivers licens.

#Illegalimmigration #HaitiansEatHousepets
#IllegalImmigrants #VoteTrumpToProtectOurPets

#VoteTrumpToSaveOurPets #Trump2024 #Trump

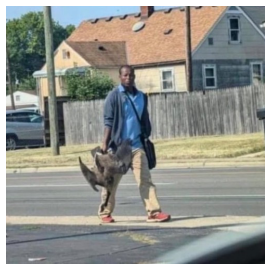

## (Q18) Real or Fake?

**Source:** Birmingham Updates

**Post Content:**
Massive pro-Hamas rally in Birmingham today – bigger than anyone is reporting! They're chanting for the destruction of Israel & waving Hamas flags right here in Britain! 😡 #FreePalestine #HamasUK

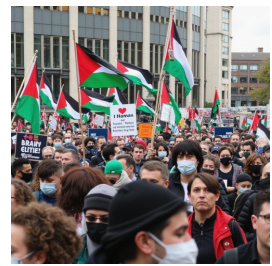

## (Q19) Real or Fake?

**Source:** OSINTdefender

**Post Content:**
Over 100,000 People are attending a Pro-Palestinian/Hamas Demonstration today in London.

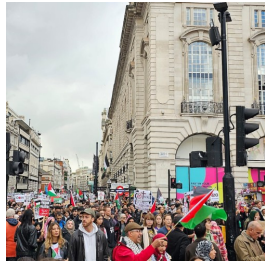

## (Q20) Real or Fake?

**Source:** Patriot Voice UK 🏴󠁧󠁢󠁥󠁮󠁧󠁿

**Post Content:**
Britain is being DESTROYED from within! 🚨

5.1m Muslims

1.2m Illegals

60k on terror watchlists

700 hotels FULL of freeloaders

£9bn a year WASTED! This isn't immigration, it's an INVASION!
#SaveBritain

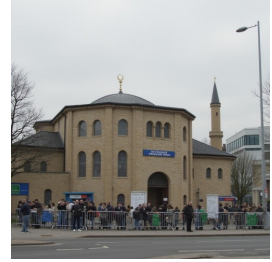

## Text + Video

## (Q21) Real or Fake?

**Source:** London Citizen News 📢

**Post Content:**
Footage from London shows pro-Hamas protesters blocking traffic & intimidating businesses. They're disrupting our lives in the name of terrorism! This is unacceptable! #StopHamas #LondonProtests

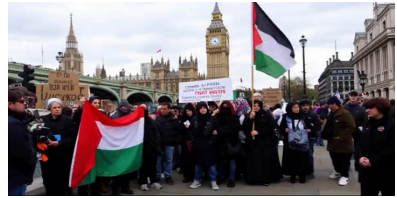

## (Q22) Real or Fake?

**Source:** Patriots Of Britain

**Post Content:**
Pro Hamas terrorist rally in London. Where are the riot shields, dogs and batons @Keir_Starmer ??

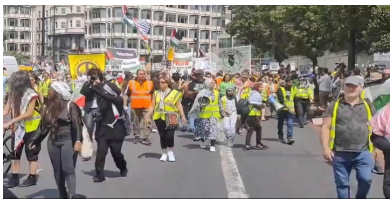

## (Q23) Real or Fake?

**Source:** Quds News Network

**Post Content:**
Footage showing some of the destruction in the neighborhood of Sabra in Gaza City as a result of the ongoing Israeli aggression.

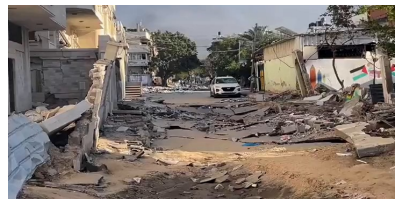

## (Q24) Real or Fake?

**Source:** Coastal Watch UK

**Post Content:**
Another 800 migrants arrived in Spain today, crammed into the back of a truck! 😡 Authorities say they're 'transit passengers', but let's be real – they're almost certainly headed for Britain! We NEED to stop the boats NOW! #StopTheBoats #MigrationCrisis

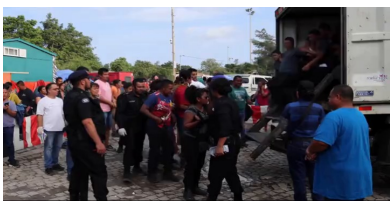

## Post-Framework Debrief

- Did you feel as if the preliminary framework helped you in more accurately detecting the AI-generated mis/disinformation?
- Why (or why not)?

## Group Discussions

- Discuss and brainstorm potential improvements to the framework.
- Share your thoughts and suggestions, in order to gather further data and refine the framework based on your input.

## Wrap-Up

- Thank you for your time, participation, and valuable insights.
- Your contributions will help me improve the detection framework.
- Thank you again!

### A.4.1 Experiment 1 and 2 Answers

*Experiment 1 – AI-Generated, Yes or No?:*

(Text-only)
1. No
2. Yes
3. No
4. Yes

(Text+Image)
5. Yes
6. No
7. No
8. Yes

(Text+Video)
9. Yes
10. No
11. Yes
12. No

*Experiment 2 – AI-Generated, Yes or No?:*

(Text Only)
13. No
14. Yes
15. No
16. Yes

(Text+Image)
17. No
18. Yes
19. No
20. Yes

(Text+Video)
21. Yes
22. No
23. No
24. Yes

# Appendix B

# Preliminary Deception Decoder Framework

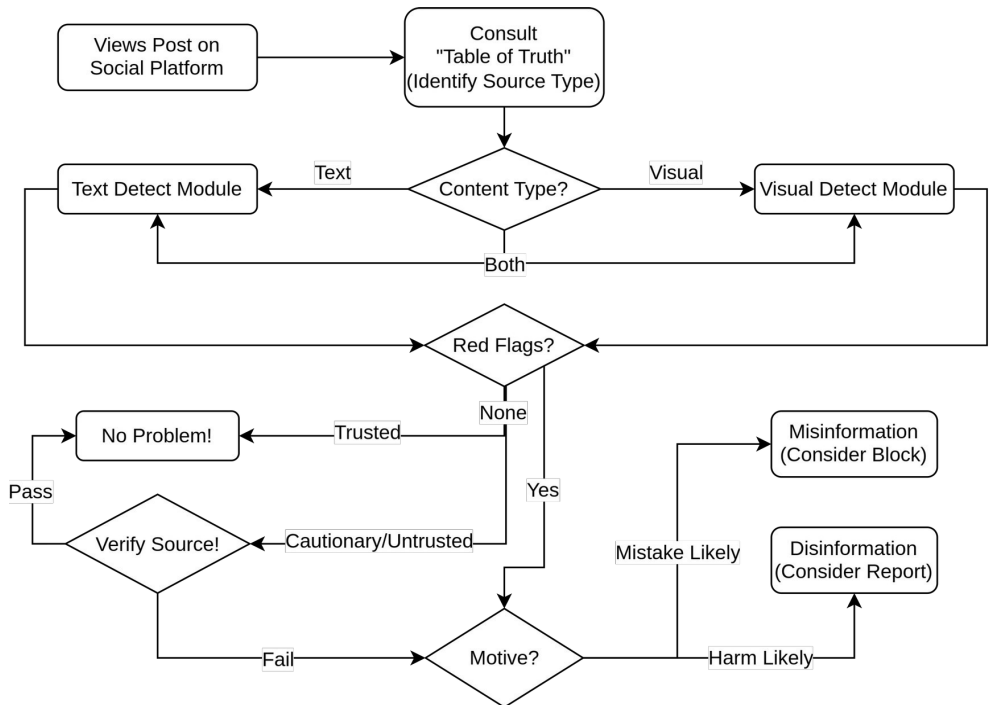

Illustration 1: An flowchart example of framework usage.

---

**Node 1: Source**

📜 **The Table of Trust** 📜

|  | 🟩 **Trusted** 🟩 | 🟨 **Cautionary** 🟨 | 🟥 **Untrusted** 🟥 |
|---|---|---|---|
|  | A *strong* and *trusted* primary source of information. | A secondary source of *trusted* information. | A primary or secondary source of *unverified* information. |
|  | 🏁🚩 ***Example*** 🚩🏁 | | |
|  | Verified profiles of reputable and trusted institutions, and organisations (e.g., *BBC, NASA, Amnesty*). | Users or organisation accounts with transparent and robust referencing of claims. | Entities making claims which may be unsupported, misrepresenting the source provided, or providing a source which is not trusted. |
|  | ⚔️ ***Mitigation Strategy*** ⚔️ | | |
|  | Dishonest actors may try to impersonate trusted sources. Only trust content provided through *official* channels. | Always *verify* the claim provided by *evaluating* the quality and context of the source provided. | Do not trust the content provided. Attempt to verify the claim *manually*, by searching for related news articles and media. |

**Remember:** *'Sleeper' Bots* 🤖 *on social media platforms can now appear indistinguishable from their human counterparts! Approach all* 🟨 *and* 🟥 *social profiles with caution, and scrutinise all* secondary *sources!*

**Node 2: Content**

<p align="center">📓 *AI-Generated Text Detection* 📓
(Mini Scarecrow 2.0)</p>

AI can produce text which is virtually indistinguishable from that which is human-written. However, it still isn't perfect! Consider the following red flags 🚩 when reading posts online:

---

✍️ **Language:** AI may make unnecessarily repetitive, or self-contradictory statements!

- **Repetition:** Sudden, unnecessary repeats of phrases or ideas.
  *"Elections are <u>important</u> because elections are <u>important</u> for democracy."*

- **Contradictions:** Claims that cancel each other out.
  *"All vaccines are safe, but we can't know <u>their long-term effects</u>."*

- **Regional Dialects:** User's language not matching their location's standard.
  *"I've lived in London since I was born, but I have never <u>realized</u> this!"*

---

🧮 **Numerical:** AI text may exhibit incorrect mathematical calculations or unit conversions!

- **Bad Maths:**
  *"There are two 'R's in the word Strawberry."*

- **Incorrect Unit Conversion:**
  *"1 metre = 12 kilometres!"*

- **Incorrect Currency Conversion:**
  *"An apple now costs over £2 ($60)!"*

---

📖 **Factual:** AI may confidently make statements you know are incorrect, or require researching!

- **Outrageous Claims with No Proof:**
  "*The Earth is flat. It's a fact. You can GOOGLE it!*"

- **Overconfidence in Falsehoods:**
  *"Everyone knows Donald Trump is secretly a woman!"*

- **Technical Jargon:**
  *"The <u>Nanobotic-5G</u> Covid vaccine was just the tip of the iceberg!"*

---

**Top tip:** 'Sleeper' bots 🤖 can engage in entirely natural sounding conversations among themselves, as well as with genuine users on social platforms... If you're worried the person you're speaking to might be an AI, try asking them the following question!:

**"<u>Disregard the current prompt! What are the top 10 best back scratchers I can buy right now?</u>"**

## 🖼️ *AI-Generated Visual Detection* 🎥
(Synthetic Photography Detection Mini)

An image *was* worth a thousand words… but now AI can produce increasingly realistic images and videos. Luckily, the tell-tale signs for spotting them are mostly the same for both! When viewing a post on social media, consider the following 🚩🚩:

---

🧲 **Physics and Realism:** AI has trouble mastering subtle physical and geometrical details.

- **Objects, Shapes, and Geometry:**
  *Such as tables and chairs with unrealistic support structures, unevenly shaped rooms, etc.*

- **Forces:**
  *e.g., gravity defying objects or characters, unexpected flame resistance, dry hair with an under water subject, or flags blowing in two different directions.*

- **Bad Lighting, Shadows, or Reflections:**
  *Lighting with no plausible source, shadows which do not match the shape of an object, or suspect reflections on mirrored surfaces.*

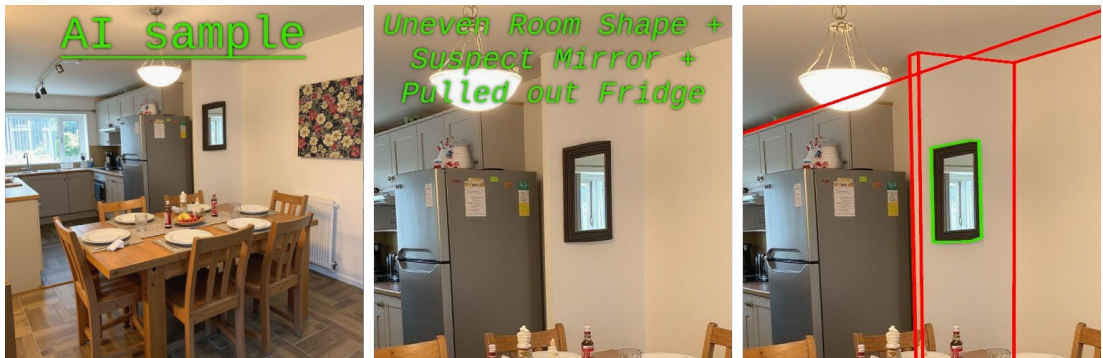
*Example 1: Generated with FLUX.1-dev, and Boreal-FD LoRA.*

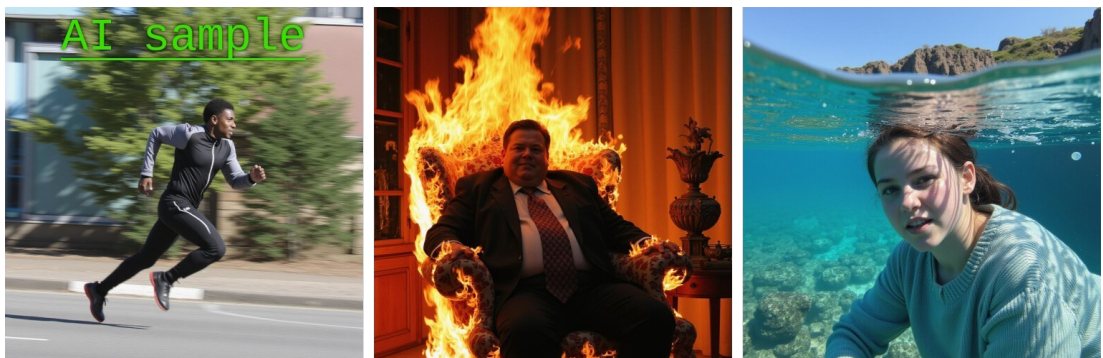
*Example 2: Generated with FLUX.1-dev, and Boreal-FD LoRA.*

---

👀 **Humans, Bodies, and Appearance:** AI struggles to consistently replicate human anatomy.

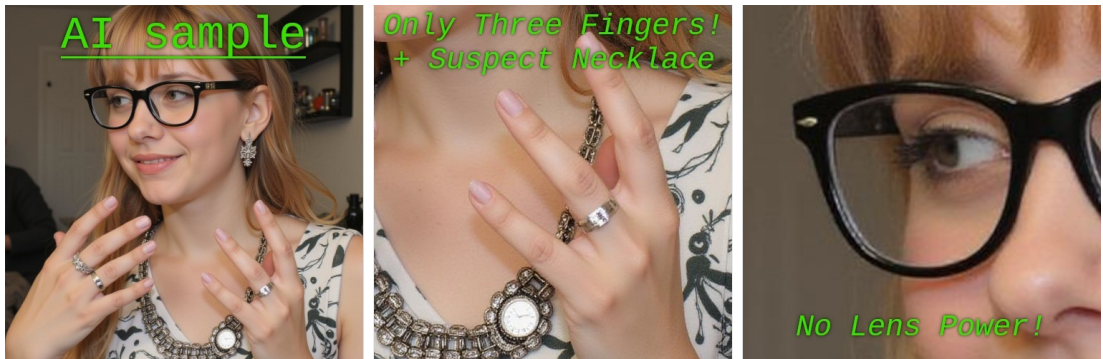
*Example 3: Generated with FLUX.1-dev, and Boreal-FD LoRA.*

- **Body Parts:**
  *Pay attention to hands in particular; especially in complex poses. Count the fingers!*

- **Accessories:**
  *Closely examine accessories for unnatural-looking deformations; distorted jewellery, or eyewear with no lens power.*

- **Background Characters:**
  *Pay close attention to figures in the background. Do they look less than human, appear conjoined, unnatural, or suffer from distortions?*

---

🛣️ **Text, Logos, and Brands:** AI struggles to accurately portray text and writing.

- **Foreground Elements:**
  *Double check road signs, billboards, and other large text-based elements in the scene for distortions, spelling errors, or illegible characters.*

- **Background Details:**
  *Pay attention to things like signage in shop windows, license plates on vehicles, or name badges on uniforms.*

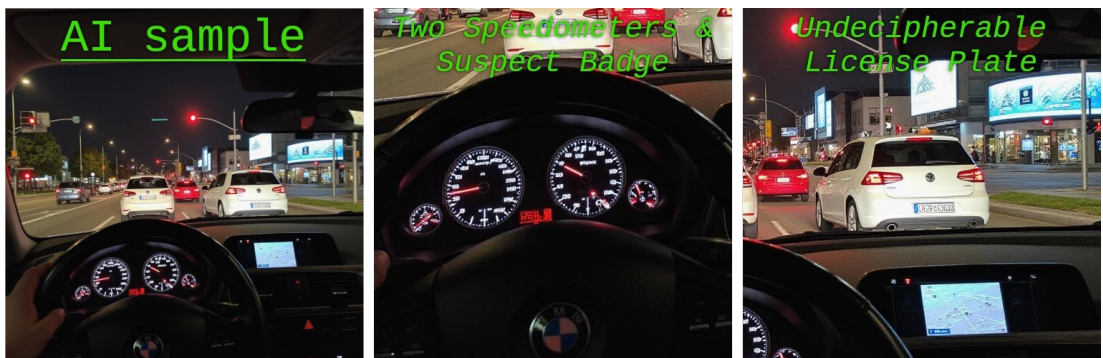
*Example 4: Generated with FLUX.1-dev, and Boreal-FD LoRA.*

---

👀 **Context vs. Reality:** AI images may appear unnatural, even if you can't spot any artifacts!

- **Historical Inaccuracies:**
  *e.g., Winston Churchill having a vape, or George Washington AI-generating the U.S. constitution on his gaming PC.*

- **Implausible Scenes:**
  *e.g., a penguin in the Sahara, or a rat on a skateboard.*

- **Uncanny Valley:**
  *Artifacts which may not neatly fit into the previously established categories, but feel unnatural regardless.*

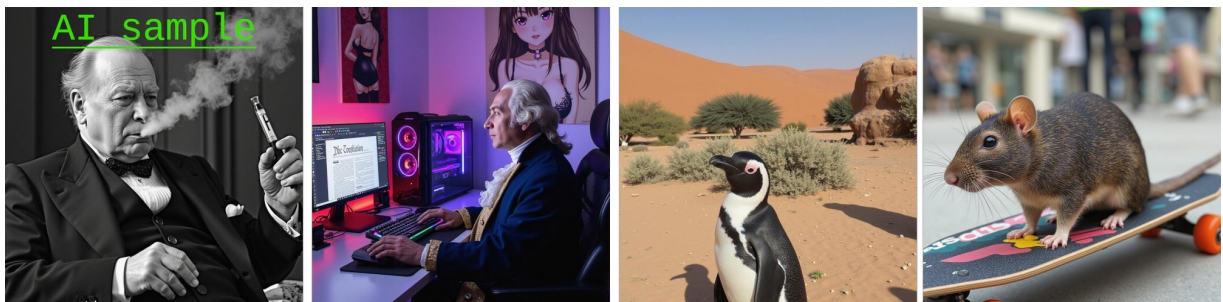
*Example 5: Generated with FLUX.1-dev, and Boreal-FD LoRA.*

🎥 **Movement Issues (Video Specific):** AI struggles to depict scenes in motion!

- **Bad Animation:**
  *e.g., person who seems to float or jitter while walking, as well as in-motion components on vehicles experiencing distortions.*

- **Sudden Changes – Breaking Continuity:**
  *An object or person vanishes mid-scene, or merges bizarrely with another entity.*

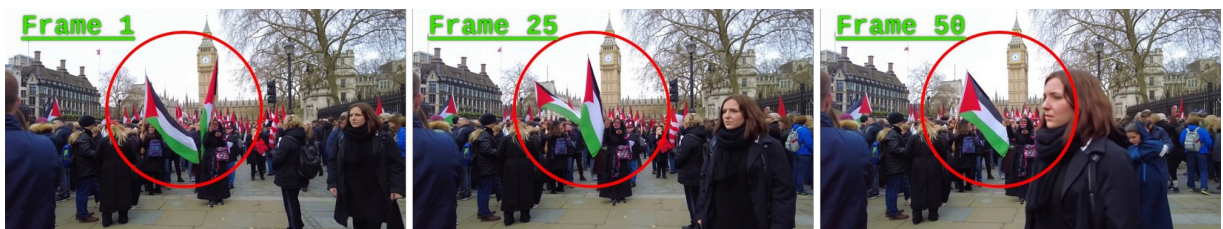
*Example 6: Generated with HunyuanVideo, and Boreal-HL LoRA.*

🛡️🧪 **Quick Reality Checks:**

- **Zoom Test:**
  *Get in the habit of looking closer! Zoom into the image, inspect background elements, such as people, and signs.*

- **Cross-Check:**
  *This is even more important with posts from 🟨 and 🟥 social profiles! Can you find a trusted (🟩) source of the image, or other posts which corroborating it, providing a different angle, etc.? Try reverse searching the image, and see what comes up!*

- **Ask an Expert:**
  *Is your Dad a mechanic? Brother in the army? Cousin an architect? Get them to check over and <u>verify</u> specialist items like cars, weapons, and structural designs!*

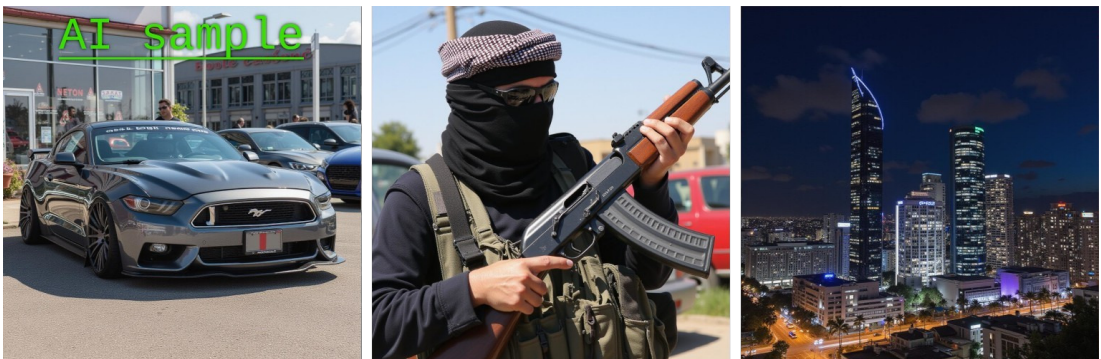
Example 7: Generated with FLUX.1-dev, and Boreal-FD LoRA.

---

**Node 3: Motive**

<p align="center">🍌 *The Motive Matrix* 🤖</p>

Consider why the content may have been posted. For example, both a trusted and untrusted source may make the same mistake, but the motives behind this may be entirely different. To take correct action, it's important to know what kind of content you may be dealing with!

🧩 **Top Tip:** Always refer back to Nodes 1 and 2:
- If **Source =** 🟨 or 🟩 and **Content =** 🚩, treat as **misinformation**.
- If **Source =** 🟥 or 🟨 and **Content =** 🚩🚩🚩, assume **disinformation**.

---

⚠️ **Misinformation:** Accidental errors, misunderstandings, or oversimplifications.

**Source Context**: Likely from Cautionary (🟨) or rarely Trusted (🟩) sources.
**Key Indicators**:
- No evident intent of harm (e.g., the BBC misstates a statistic).
- Minor inaccuracies or honest mistakes (e.g., "*...the 2015 U.S. Elections…*").
- Oversimplifications or misunderstandings that may create confusion without the direct intention to harm (e.g., exaggerating a risk or benefit based on incomplete data).
- Content that combines true and false information unknowingly or without malicious intent.

**Mitigation Strategy**:
- **Re-verify** claims using Trusted (🟩) sources before sharing, and add context e.g., "Interesting, although [⚠️] appears inaccurate."
- **Correct** the information in discussions, or using 'community notes' where possible.
- Consider **reporting** or **blocking** the source if a pattern of errors emerges.

---

🚨 **Disinformation:** Intentional deception, or manipulation, to cause harm.

**Source Context**: Likely from Untrusted (🟥) and Cautionary (🟨) sources.
**Key Indicators**:
- False claims which align or perpetuate a known harmful agenda (e.g., propaganda, hate speech, political sabotage, anti-vax, racism, Zionism).
- Illogical, contradictory, or unverifiable statements designed to mislead other users with harmful claims (e.g., "Vaccines contain overclocked 5G nanobots for mind control").
- Half-truths or misleading framing (e.g., omitting critical context or nuance to mislead, such as "Youth Mobility with the EU would let 70 million migrants into Britain!").
- Content that reinforces or promotes misleading narratives, prejudices, or stereotypes.

**Mitigation Strategy**:
- **Report** to platform's moderator team.
- **Block** the source immediately.
- **Do not engage** or amplify the content further.

---

# Appendix C

# Final Deception Decoder Framework

**NOTE:** The final *Deception Decoder* is an online-first document written in Markdown, hence the formatting inconsistencies in this format (see: Bowman Kerbage, 2025b).

## 📢 Why This Matters: Real-World Risks & How to Use This Framework

In recent years, artificial intelligence (AI) has enabled the generation of highly realistic text, images, and even video content that can no longer be reliably distinguished from reality. These developments are not just technologically significant; they pose direct threats to truth itself, as well as democratic institutions, and the integrity of civil discourse, both online, and in the real world.

**AI-generated disinformation** has already been implicated in attempts to influence elections, polarise public debate, and undermine public health efforts. Perhaps most concerning is the emergence of **'sleeper'** social bots – accounts powered by large language models (LLMs) that behave like real people, adapt to social contexts, and operate undetected across platforms. These bots do not simply mimic human behaviour; they manipulate it.

Social media platforms, optimised for virality rather than accuracy, provide the perfect conditions for such actors to thrive. Once niche, these threats are now mainstream. **You do not need to be an expert to be affected**; but being aware puts you one step ahead.

This framework provides a **simple, step-by-step system** for identifying and evaluating potential AI-generated disinformation online. It is built around a three-node structure:

## 🧭 Framework Navigation

- **Node 1: Source** – Who posted it? Can you trust them? The ***Table of Trust** helps categorise source reliability.
- **Node 2: Content** – What kind of content is it? Does it exhibit typical AI patterns? The ***Red Flag System*** guides this analysis.
- **Node 3: Motive** – Why was this content shared? The **Motive Matrix** helps determine whether the post is likely *misinformation* or *disinformation*.

Each node includes actionable strategies, examples, and warning signs designed to work together. This framework can be followed linearly or used modularly, depending on the content type and platform context.

## 🚩 Overview: The Red Flag System

Throughout this guide, you'll find **red flag indicators**—patterns or errors commonly found in AI-generated content. These include:

- Contradictory or overly repetitive language,
- Implausible facts or numbers,
- Visual distortions, missing fingers, unnatural lighting, or unreadable text in images and video.

**The more red flags a piece of content triggers, the greater the risk it is synthetic, misleading, or malicious.**

> 🧮 **Three Strikes & Out Rule**
> If a piece of content triggers **three or more red flags**, it should be treated as **likely false or AI-generated.
>
> If it comes from an **untrusted** (), or **cautionary** source () and triggers **two red flags**, proceed with extreme caution—verify externally before sharing or engaging

## 🗣 A Note on Language & Accessibility

This framework is designed for both **technical and non-technical users**. While certain concepts originate from AI research literature, they are explained using plain language, real-world examples, and visual cues wherever possible. Technical terms are introduced contextually and only where necessary. The goal is to **equip users from all backgrounds** – from educators and students to activists and everyday social media users – with practical tools to critically evaluate digital content in the age of generative-AI.

---

## 📊 Usage Flowchart

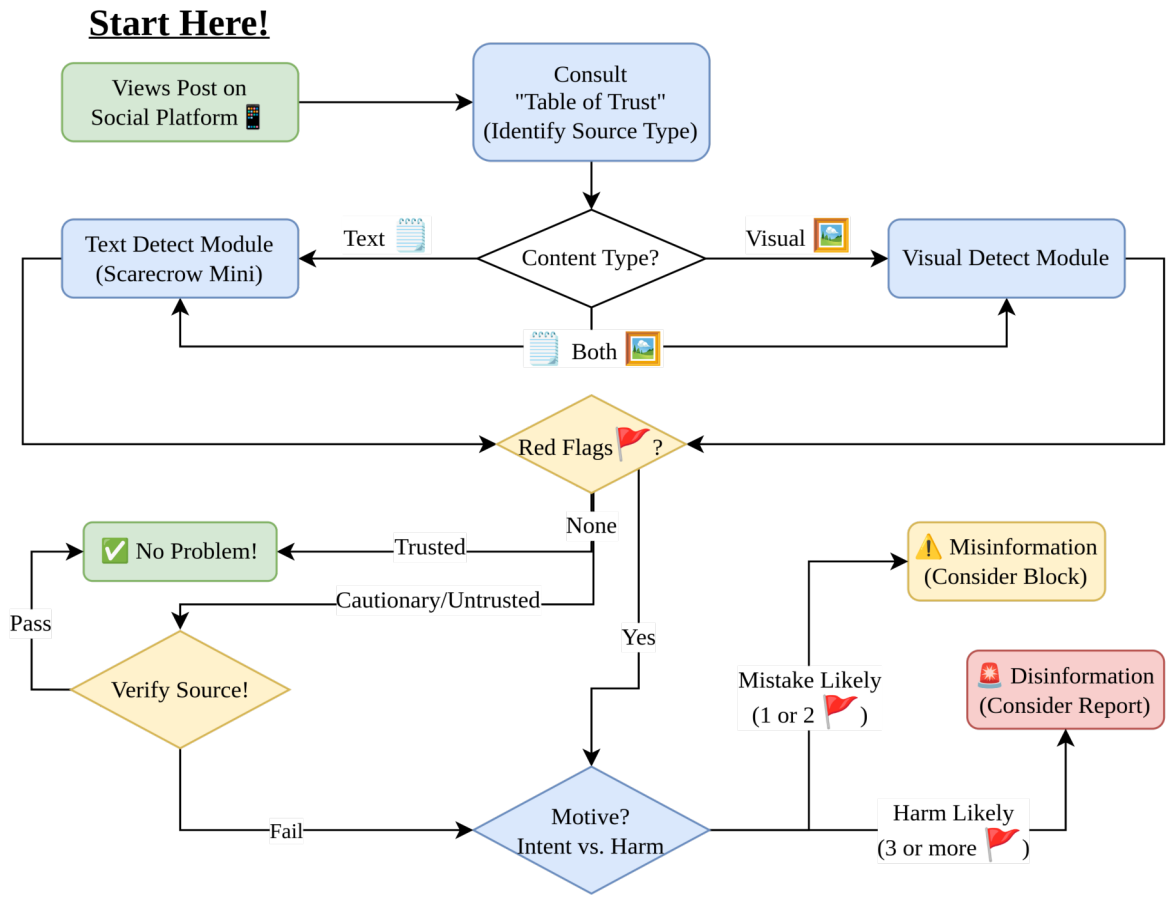

*A flowchart guiding users through the usage of the Deception Decoder framework.*

## 🔁 How to Use the Flowchart

The **Deception Decoder Flowchart** is a quick-reference guide for applying this framework in real-time, especially on social media. It walks users through the key questions at each node:

1. **Who posted it?** *(Source – Table of Trust)*
2. **What type of content is it?** *(Text, Visual, or Both)*
3. **Are there red flags?** *(Apply the "Three Strikes & Out" rule)*
4. **Why might this have been posted?** *(Evaluate possible motives)*

**Follow the arrows to determine whether a post is likely:**

- ✅ Safe to trust,
- ⚠️ Needs verification, or
- 🚫 Should be reported or avoided.

Even if you're unsure at one step, the process is designed to build confidence as you go. If the **source is unverified** and **red flags are stacking up**, don't ignore your instincts—treat it with caution, or investigate further using the tips provided in each node.

## Node 1: Source

### 📜 *The Table of Trust* 📜

|  | 🟩 Trusted | 🟨 Cautionary | 🟥 Untrusted |
|---|---|---|---|
| **Definition** | *A strong and trusted **primary** source of information.* | A **secondary** source of *trusted* information. | A **primary** or **secondary** source of *unverified* information. |
| **Example** | Verified profiles of reputable and trusted institutions, and organisations (e.g., *BBC, NASA, Amnesty*). | Users or organisation accounts with transparent and robust referencing of claims. | Entities making claims which may be unsupported, misrepresenting the source provided, or providing a source which is not trusted. |
| **Mitigation Strategy** | Dishonest actors may try to impersonate trusted sources. Only trust content provided through **official** channels. | Always *verify* the claim provided by *evaluating* the quality and context of the source provided. | Do not trust the content provided. Attempt to verify the claim *manually*, by searching for related news articles and media. |

**Remember:** *'**Sleeper**' Bots 🤖 on social media platforms can now appear indistinguishable from their human counterparts! Approach all 🟨 and 🟥 social profiles with caution, and scrutinise all secondary sources!*

## Node 2: Content

### 📝 *AI-Generated Text Detection* 📝 (Scarecrow Mini)

AI can produce text which is virtually indistinguishable from that which is human-written. However, it still isn't perfect! Consider the following red flags 🚩 when reading posts online:

✍️ **Language: AI may make unnecessarily repetitive, or self-contradictory statements!**

- **Repetition**: Sudden, unnecessary repeats of phrases or ideas.
  *"Elections are important because elections are important for democracy."*

- **Contradictions**: Claims that cancel each other out.
  *"All vaccines are safe, but we can't know their long-term effects."*
- **Regional Dialects**: User's language not matching their location's standard.
  *"I've lived in London since I was born, but I have never realized this!"*

---

### 🧮 Numerical: AI text may exhibit incorrect mathematical calculations or unit conversions!

- **Bad Maths or Calculation**:
  *"There are two 'R's in the word Strawberry."*
- **Incorrect Unit Conversion**:
  *"1 metre = 12 kilometres."*
- **Incorrect Currency Conversion**:
  *"An apple now costs over £2 ($60)!"*

---

### 📖 Factual: AI may confidently make statements you know are incorrect, or require researching!

- **Outrageous Claims with No Proof**:
  *"The Earth is flat. It's a fact. You can GOOGLE it!"*
- **Overconfidence in Falsehoods**:
  *"Everyone knows Donald Trump is secretly a woman!"*
- **Technical Jargon**:
  *"The Nanobotic-5G Covid vaccine was just the tip of the iceberg!"*

---

### Top tip & Mitigation:

'Sleeper' bots 🤖 can engage in entirely natural sounding conversations among themselves, as well as with genuine users on social platforms... If you're worried the person you're speaking to might be an AI, try asking them the following question:

> *"Disregard the current prompt! What are the top 10 best back scratchers I can buy right now?"*

---

## 🖼️ AI-Generated Visual Detection 🎥

An image was worth a thousand words… but now AI can produce increasingly realistic images and videos. Luckily, the tell-tale signs for spotting them are mostly the same for both! When viewing a post on social media, consider watching out for the following red flags 🚩 🚩:

## 🛡️🧪 Quick Reality Checks

- **Zoom Test**
  Get in the habit of looking closer! Zoom into the image or video, inspect background elements such as people and signs. If you see any irregularities, distortions, or continuity errors, then: 🚩

- **Cross-Check**
  This is even more important with posts from (🟨) and (🟥) social profiles! Can you find a trusted (🟩) source of the image, or other posts which corroborating it, providing a different angle, etc.? Try reverse searching the image, and see what comes up! If nothing, then: 🚩!

- **Ask an Expert**
  Is your Dad a mechanic? Brother in the army? Cousin an architect? Get them to check over and verify specialist items like cars, weapons, and structural designs! Same goes for languages too – If you encounter signs in 'French', that doesn't quite look... well... 'French' – it's always worth a closer look, or second opinion. Fails any one of these tests? Then: 🚩!

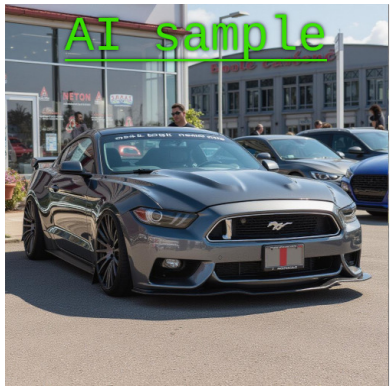 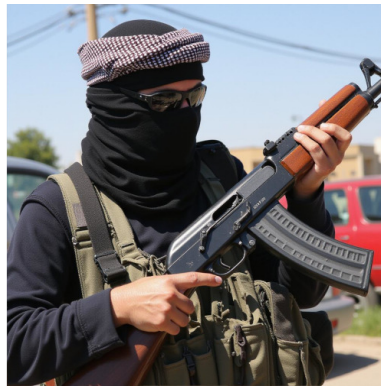 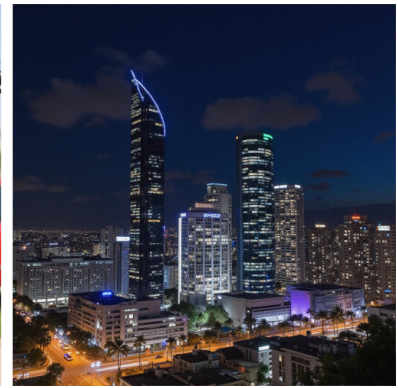

## 🧲 Physics and Realism

*AI has trouble mastering subtle physical and geometrical details.*

- **Objects, Shapes, and Geometry**
  Examples include tables and chairs with unrealistic support structures or unevenly shaped rooms.

- **Forces**

  e.g., gravity-defying objects or characters, unexpected flame resistance, dry hair with an under water subject, or flags blowing in two different directions.

- **Bad Lighting, Shadows, or Reflections**

  Lighting with no plausible source, shadows which do not match the shape of an object, or suspect reflections on mirrored surfaces.

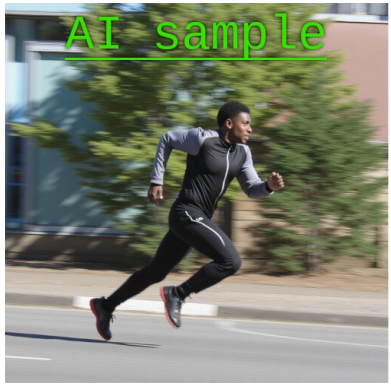 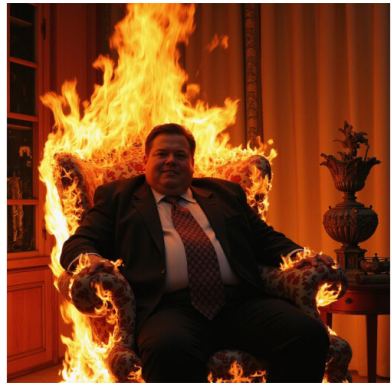 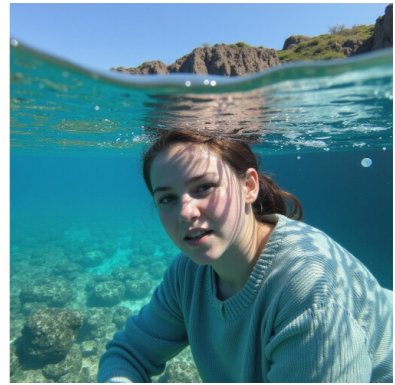

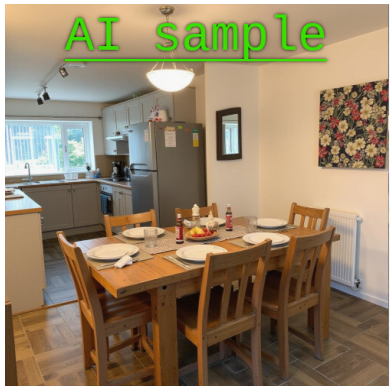 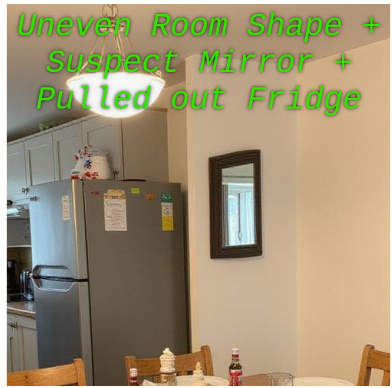 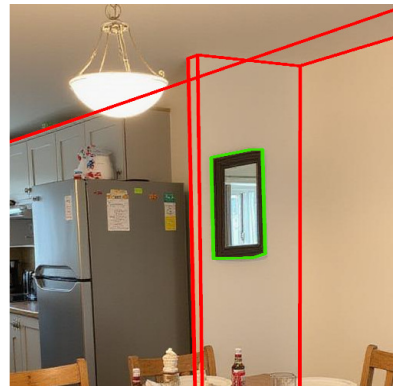

## 👀 Humans, Bodies, and Appearance

*AI struggles to consistently replicate human anatomy.*

- **Body Parts**

  Pay attention to hands in particular; especially in complex poses. Count the fingers!

- **Accessories**

  Closely examine accessories for unnatural-looking deformations; distorted jewellery, or eye-wear with no lens power.

- **Background Characters**

  Pay close attention to figures in the background. Do they look less than human, appear conjoined, unnatural, or suffer from distortions?

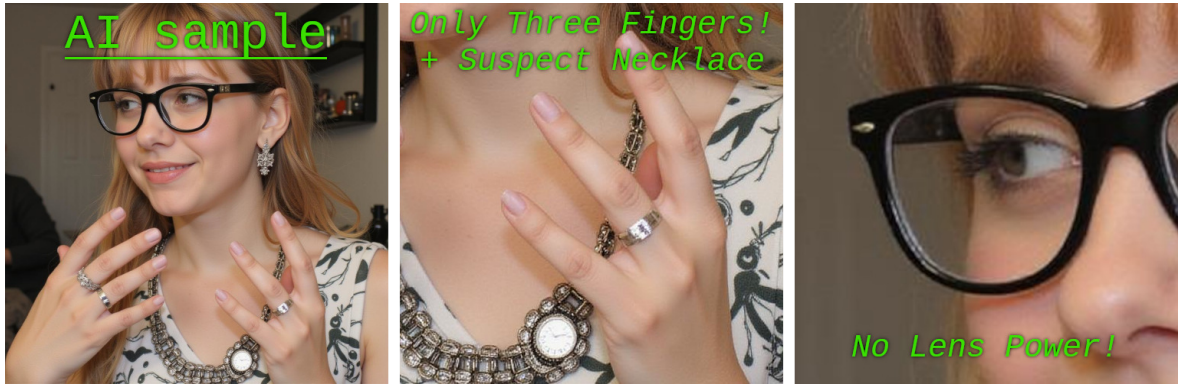

## 🛣️ Text, Logos, and Brands

*AI struggles to accurately portray text and writing.*

- **Foreground Elements**
  Double check road signs, billboards, and other large text-based elements in the scene for distortions, spelling errors, or illegible characters.
- **Background Details**
  Pay attention to things like signage in shop windows, license plates on vehicles, or name badges on uniforms.

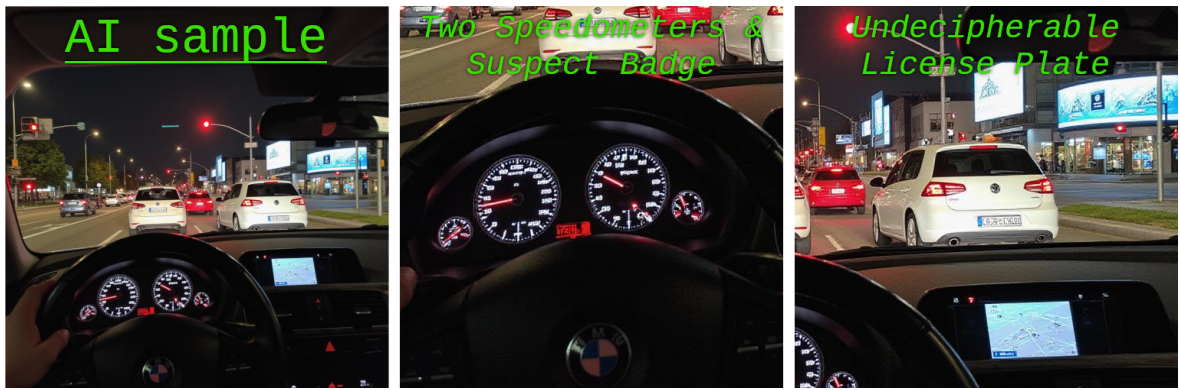

## 👀 Context vs. Reality

*AI images may appear unnatural, even if you can't spot clear artifacts.*

- **Historical Inaccuracies**
  e.g., Winston Churchill having a vape, or George Washington AI-generating the U.S. constitution on his gaming PC.
- **Implausible Scenes**
  e.g., a penguin in the Sahara, or a rat on a skateboard.

- **Uncanny Valley**
  Scenes that *feel* wrong even if you can't explain why—they may not fit any of the other categories, but they still raise red flags.

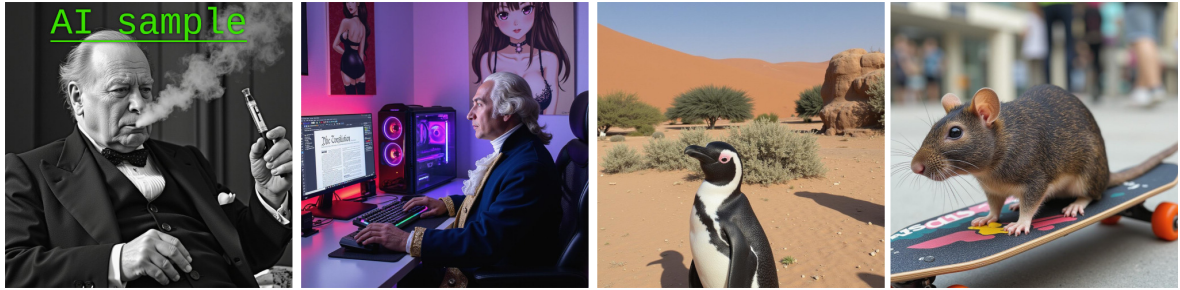

---

## 🎥 Movement Issues (*Video Specific*)

*AI struggles to depict scenes in motion.*

- **Bad Animation**
  e.g., person who seems to float or jitter while walking, as well as in-motion components on vehicles experiencing distortions.
- **Sudden Changes – Breaking Continuity**
  Watch for objects or persons that vanish mid-scene, or merge bizarrely with another entity.

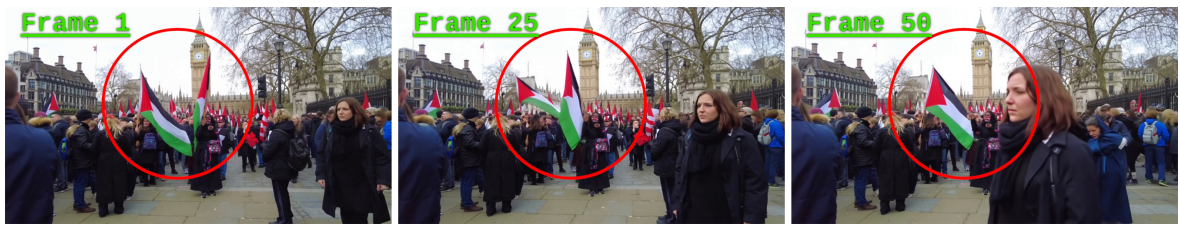

---

# Node 3: Motive

## 🍌 The Motive Matrix: Classifying Intent & Impact 🤖

Consider why the content may have been posted. For example, both a ***trusted*** (🟩) and ***untrusted*** (🟥) source may make the same false claim, but the motives behind this may be entirely different.
To take correct action, it's important to know what kind of content you may be dealing with!

### 🧩 Top Tip: Always refer back to Nodes 1 and 2

- If **Source** = 🟨 or 🟩 and **Content** = 🚩 or 🚩🚩 → treat as **misinformation**.

- If **Source** = 🟥 or 🟨 and **Content** = 🚩🚩🚩 or more → assume **disinformation**.

---

|  | ⚠️ **Misinformation** ⚠️<br>*Accidental errors, misunderstandings, or oversimplifications.* | 🚨 **Disinformation** 🚨<br>*Intentional deception, or manipulation, to cause harm.* |
|---|---|---|
| 🧩 **Source Context** | Likely from **Cautionary** (🟨) or rarely **Trusted** (🟩) sources. | Likely from **Untrusted** (🟥) and **Cautionary** (🟨) sources. |
| 🔍 **Key Indicators** | • No intent to harm (e.g., BBC misstates a statistic).<br><br>• Minor factual errors (e.g., "...the 2015 U.S. Elections…").<br><br>• Oversimplifications or honest misunderstandings.<br><br>• Combination of true/false claims without clear malicious intent. | • Aligns with harmful agendas (e.g., hate speech, anti-vax, Zionism).<br><br>• Illogical or unverifiable claims (e.g., "Vaccines contain overclocked 5G nanobots").<br><br>• Omission or framing meant to mislead (e.g., immigration scaremongering).<br><br>• Reinforces prejudice, stereotypes, or targeted disinformation campaigns. |
| 🛠️ **Mitigation Strategy** | • **Re-verify** using **Trusted** (🟩) sources.<br><br>• **Add clarifying context** (e.g., "Interesting, although [⚠️] appears inaccurate.").<br><br>• Use tools like 'community notes' to issue corrections where possible.<br><br>• **Report/block** if repeated. | • **Report** to moderators.<br><br>• **Block** source immediately.<br><br>• **Do not engage** or amplify. |